\documentclass[11pt, a4paper, onecolumn, copyright]{AweAI}

\usepackage[authoryear, sort&compress, round]{natbib}
\usepackage[utf8]{inputenc}
\usepackage[T1]{fontenc}
\usepackage{hyperref}
\usepackage{url}
\usepackage{booktabs}
\usepackage{amsfonts}
\usepackage{nicefrac}
\usepackage{microtype}
\usepackage{amsmath,amssymb,amsthm}
\usepackage{bm}
\usepackage{multirow}
\usepackage{graphicx}
\usepackage{float}
\usepackage{array}
\usepackage{tabularx}
\usepackage{ragged2e}
\usepackage{makecell}
\usepackage{balance}
\usepackage[normalem]{ulem}
\usepackage[table]{xcolor}
\usepackage{colortbl}
\usepackage{algorithm}
\usepackage{algorithmicx}
\usepackage[noend]{algpseudocode}
\usepackage{enumitem}
\usepackage[most]{tcolorbox}
\usepackage{marvosym}

\usepackage{adjustbox}
\usepackage{tikz}
\usetikzlibrary{shapes,arrows.meta,positioning,fit,matrix}
\usepackage[edges]{forest}

\definecolor{lightgray}{RGB}{215,215,215}
\definecolor{myred}{RGB}{210,109,91}
\definecolor{mygreen}{HTML}{F28705}
\useunder{\uline}{\ul}{}
\newcolumntype{L}{>{\RaggedRight\arraybackslash}X}

\newcommand{\ie}{\emph{i.e., }}
\newcommand{\eg}{\emph{e.g., }}

\newcommand{\cf}{\emph{cf. }}

\newlength\myindent
\floatname{algorithm}{Algorithm}

\theoremstyle{definition}
\newtheorem{definition}{Definition}[section]
\theoremstyle{remark}

\title{Autonomous Information Seeking: A Roadmap for Agentic Recommender Systems}

\author{%
  {Xinyu Lin$^1$~Yashar Deldjoo$^2$\textsuperscript{\Letter}~Sunhao Dai$^3$~Honghui Bao$^1$~Xiaopeng Ye$^3$~Fatemeh Nazary$^2$~
  Wenjie Wang$^4$~Tommaso Di Noia$^2$~Jun Xu$^3$~Tat-Seng Chua$^1$}\\
  $^1$National University of Singapore~~$^2$Polytechnic University of Bari\\
  $^3$Renmin University of China~~$^4$University of Science and Technology of China\\
  \texttt{xylin1028@gmail.com, deldjooy@acm.org}
}

\renewcommand{\today}{}
\makeatletter
\renewcommand{\abscontent}{%
    \noindent
    \parbox{\dimexpr\linewidth}{\absfont \theabstract}%
    \vskip1em
    \noindent\textbf{Keywords:} Agentic Recommender System, Level of Autonomy, LLM Agent, Agent-assisted Recommender, Agent as Recommender, Agent as User Simulator\\
    \vspace{0.4em}
    \hrule
    \vspace{0.3em}
    {\small
    \textsuperscript{\Letter} Corresponding author.
    }
}
\fancypagestyle{firststyle}{
    \fancyhead[R]{%
        \ifdefined\paperurl
        \if\relax\the\paperurl\relax \else
            \href{\the\paperurl}{\urlheaderfont \itshape \the\paperurl}\\ \fi
        \else
        \fi
        {\footerfont\itshape\monthyeardate\today\hspace{8pt}}%
        \ifthenelse{\boolean{confidential}}{\\ \footerfont \internalonly}{}
    }
    \fancyhead[C]{}
    \fancyfoot[L]{
    \footerfont\itshape
        \ifdefined\correspondingauthor
        \if\relax\detokenize\expandafter{\the\correspondingauthor}\relax
      \else
        \footerfont\itshape {Emails: \the\correspondingauthor\\}%
      \fi
    \fi
    }
    \fancyfoot[R]{
        \ifthenelse{\boolean{confidential}}{
        \ifdefined\reportnumber
        \if\relax\the\reportnumber\relax
        \else \footerfont\itshape  {\footerfont \thepa{} Technical Report \the\reportnumber} \fi
        \else \fi
        }{\footerfont\bfseries\relax}
    }
    \fancyfoot[C]{
        \footerfont\bfseries\relax}
}
\makeatother

\begin{abstract}
The rapid integration of large language model-based agents into recommender systems has driven a shift from static, ranking-based pipelines toward autonomous and interactive systems that can reason, plan, and act. This survey provides a comprehensive overview of this emerging landscape by introducing a unified taxonomy grounded in the level of autonomy and three core paradigms of agentic recommender systems: agent-assisted recommendation, agent-as-recommender, and agent-as-user-simulator. The autonomy framework organizes existing methods along increasing capabilities in proactivity, context awareness, interaction flexibility, and adaptivity. Building on this framework, the survey analyzes how each paradigm adopts different agentic architectures and how agents enhance key components such as profiles, memory, tool use, workflows, and optimization mechanisms.
We further examine evaluation methodologies for agentic recommendation, covering automated metrics, LLM-based judging, and simulation-based assessment, and discuss their limitations in capturing reasoning quality, user experience, and system behavior. 
Beyond existing evaluation protocols, we further discuss unresolved issues in evaluating agentic recommender systems, including trajectory-level assessment, agent contribution analysis, and calibration of user simulation. 
Lastly, the survey outlines open challenges in lifelong user modeling, contextual abstraction, multimodal alignment, controllability, trustworthiness, privacy, scalability, and efficiency. 
Together, these analyses establish a unified foundation for understanding the current progress of agentic recommender systems and highlight promising opportunities for developing more autonomous, reliable, and human-aligned recommendation agents.
\end{abstract}

\begin{document}

\begingroup
\makeatletter
\renewcommand{\thefootnote}{}
\renewcommand{\@makefnmark}{}
\maketitle
\makeatother
\endgroup

\section{Introduction}
\label{sec:introduction}

Recommender systems (RS) have traditionally been evaluated by how accurately they predict user preferences and rank items based on historical user--item interactions.
Classical approaches---from collaborative filtering to modern deep retrieval and ranking architectures trained on clicks, ratings, and purchases---encode past behavior into latent representations and, given a user and context, output a ranked list of candidate items.
This paradigm is highly effective for the dominant interaction pattern in today’s platforms: the system curates options and the user chooses among them (\eg selecting a movie from a ranked homepage list).

However, emerging recommendation scenarios increasingly involve \textbf{complex, multi-step user goals} that are difficult to support with a single ``one-shot'' ranked list, even when the underlying ranker is highly accurate. 
Users may want help planning a week-long vacation under constraints (budget, timing, transportation, lodging), adapting plans as external factors change (weather, delays), or coordinating a wardrobe under style, budget, and occasion constraints.
What makes these scenarios challenging is not merely preference prediction, but \textbf{constraint reasoning, iterative refinement, and decision support} under incomplete and evolving information.
Importantly, many real systems---and many works surveyed in this paper---still ultimately provide a ranked list or a shortlist of options.
The key shift is that the recommendation is increasingly produced through a \textbf{multi-step workflow} that must interpret goals, elicit constraints, consult tools or knowledge sources, and revise outputs as new information arrives.

\begin{table}[!t]
\centering
\caption{Modernized comparison of recommender system paradigms.}
\label{tab:system-compare-modern}
\vspace{-8pt}
\footnotesize
\renewcommand{\arraystretch}{1.3}
\setlength{\tabcolsep}{5pt}
\rowcolors{2}{gray!10}{white}
\begin{tabularx}{\textwidth}{@{}>{\bfseries}p{3.0cm} L L L@{}}
\rowcolor{gray!20}
\textbf{Dimension} & \textbf{Classical RS} & \textbf{LLM-based RS (generative / prompt-based)} & \textbf{Agentic RS} \\
\toprule
\textbf{Recommendation Goal}
& {Personalized ranking}: match past user--item interactions via shallow user models.
& {Personalized ranking \& language generation}: infer preferences via prompts and generate textual outputs.
& \textbf{\emph{Goal-oriented decision support}}: produce recommendations under constraints via multi-step reasoning and actions. \\
\textbf{Proactivity}
& Reactive: suggest only when RS is explicitly requested.
& Mostly reactive: respond to prompts and conversational requests. 
& \textbf{\emph{Mixed-initiative potential}}: be able to ask clarifying questions, propose plans, and surface trade-offs. \\
\textbf{Context Awareness}
& Limited to behavior logs and static features.
& Uses context within the LLM’s window with optional retrieved snippets.
& \textbf{\emph{Situational \& tool-grounded}}: integrates logs, profiles, and external data sources (\eg web/APIs). \\
\textbf{Interactivity}
& Single-step recommendation or static re-ranking.
& Natural-language interaction, including single- and multi-turn interactions, but typically without explicit planning/control.
& \textbf{\emph{Multi-step workflows}}: plan/act/verify cycles with error recovery and iterative refinement. \\
\textbf{Adaptivity}
& Offline training with periodic model updates. 
& Limited in-session adaptation (\eg in-context learning). 
& \textbf{\emph{In-session and lifelong adaptation}}: profile/memory updates via reflection or feedback-driven adjustment. \\
\textbf{Memory}
& No explicit working memory beyond logs and engineered features.
& Mostly short context window in single-turn recommendation; conversation history in multi-turn recommendation.
& \textbf{\emph{Structured memory}}: working memory with optional episodic/semantic stores across steps/sessions. \\
\textbf{Tools \& Knowledge}
& Fixed catalog/indices without access to external knowledge.
& Parametric internal knowledge and optional external knowledge via retrieval. 
& \textbf{\emph{External tool use}}: invokes retrieval, rankers, search, APIs, or constraints solvers for grounded actions. \\
\textbf{Collaboration}
& Independent optimization of models with little to no interaction between distinct modules.
& Integrate LLMs with conventional models via pre-defined sequential workflow.
& \textbf{\emph{Inter-agent collaboration}}: enables dynamic collaboration between specialized agents to solve complex tasks through negotiation and feedback. \\
\bottomrule
\end{tabularx}
\end{table}

The rapid progress of foundation models (FMs) and large language models (LLMs) has opened new avenues for recommendation through natural-language interfaces and stronger semantic reasoning.
A growing body of work surveys LLM-enhanced recommendation and generative recommendation, emphasizing how LLMs can act as encoders, re-rankers, conversational interfaces, or generators of recommendations and explanations
(\eg \citet{wu2024survey,zhao2024recommender,li2024large}, and the broader Gen-RecSys view in \citet{deldjoo2024recommendation,deldjoo2024review}).
Yet, \textbf{LLM-based RS are not necessarily agentic}.
Many LLM-based systems remain predominantly reactive: they respond to prompts and in-context history but do not explicitly plan, call external tools, maintain persistent memory, or coordinate specialized modules beyond a single forward pass.
This matters because recommendation is a system problem: user modeling, retrieval, ranking, constraint handling, and evaluation often require structured control over external components and environment interactions. 
The comparisons between classical RS, LLM-based RS, and Agentic RS are summarized in Table~\ref{tab:system-compare-modern}, from seven dimensions, including recommendation goal, proactivity, context awareness, interactivity, adaptivity, memory, and tools \& knowledge. 

\paragraph{From ``models that rank'' to ``systems that pursue goals''.}
This survey focuses on \emph{agentic recommender systems} (ARS): systems that integrate one or more agents---often instantiated by LLMs---with explicit mechanisms for \emph{action selection} (\eg tool use and memory updates) inside a recommendation environment.
Concretely, an agent in ARS can (i) interpret a natural-language goal, (ii) decompose it into sub-tasks, (iii) invoke tools such as retrieval, ranking, filtering, web search, or APIs, (iv) track intermediate state (\eg constraints, candidate sets, user feedback), and (v) refine recommendations iteratively.
This level of interactivity and controllable action \emph{enables} a shift from ``models that rank'' toward ``systems that pursue goals,'' though we stress that today’s systems vary widely in autonomy and many still output ranked items rather than executing end-to-end tasks.

\subsection{A Level-of-Autonomy (LoA) lens for agentic recommender systems}
\label{subsec:loa}

\begin{figure}[!t]
    \centering
    \includegraphics[width=0.955\linewidth]{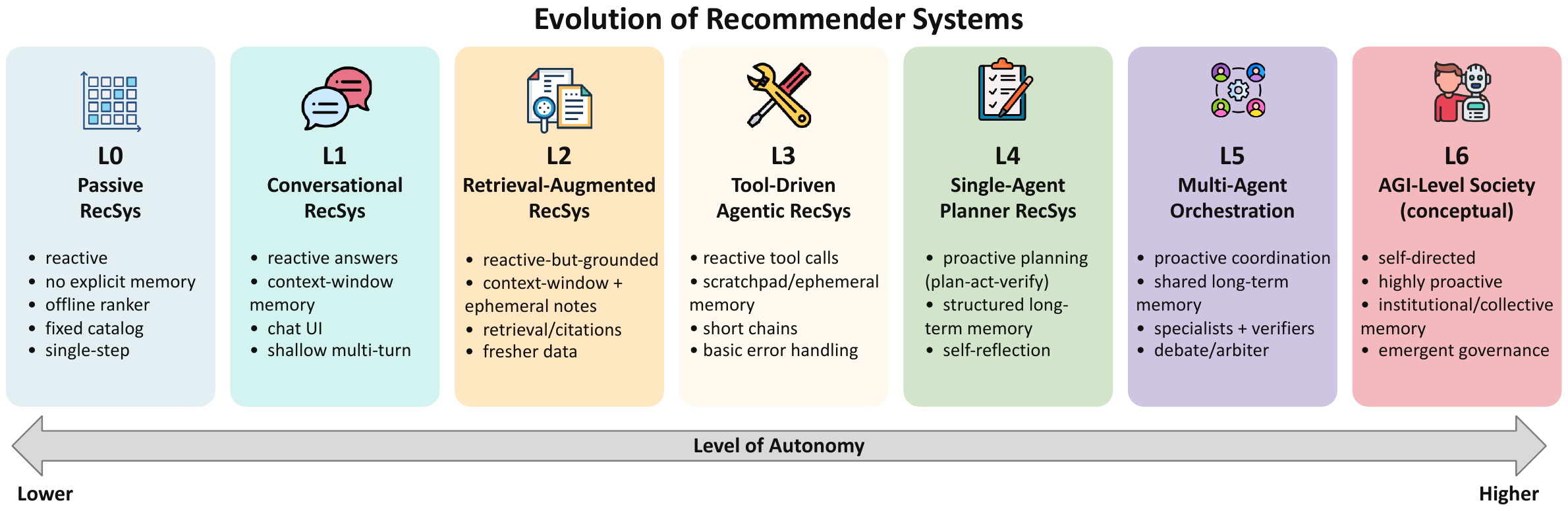}
    \caption{A Level-of-Autonomy (LoA) spectrum for recommender systems, from passive rankers (L0) to multi-agent orchestration (L5), with L6 as a conceptual ``society of agents'' endpoint.
    LoA helps connect architectural ingredients (memory, tools, orchestration) to behavioral capabilities (planning, proactivity, adaptivity) and associated risks.}
    \label{fig:loa}
\end{figure}

Because ``agentic'' is used inconsistently across the literature, we adopt a Level-of-Autonomy (LoA) perspective that situates concrete recommender architectures along a continuum from passive ranking to multi-agent orchestration.
LoA is driven by four recurring dimensions: 
(1) \textbf{Task scope and planning style}---from single-shot scoring to multi-step plan--act--verify workflows;
(2) \textbf{Context awareness and memory}---from short in-context state to persistent user/environment memory;
(3) \textbf{Interaction flexibility}---from static outputs to multi-turn, mixed-initiative, and multi-agent dialogue;
(4) \textbf{Adaptivity}---from static inference to systems that update profiles, reflect, or learn from feedback. 
Figure~\ref{fig:loa} illustrates an L0--L6 spectrum. In this survey, we focus on L2--L5, where concrete agentic architectures exist today and where autonomy-related design choices (memory, tools, orchestration, verification) materially affect recommendation quality and risk. 

In this survey, we further classify existing ARS work into \textbf{three paradigms}, \ie agent-assisted recommenders, agent-as-recommenders, and agent-as-simulators. The three paradigms are aligned with the LoA: 
agent-assisted recommenders often clusters around L2--L4, while agent-as-recommenders and simulators frequently require L4--L5 capabilities such as long-term memory, reflection, and coordination. We detail the dual taxonomy in \S~\ref{sec:foundations-taxonomy}. 

\subsection{Task landscape and survey scope}
\label{subsec:task-scope}

Agentic recommender systems span a broader task space than classical ``top-$N$ ranking.'' 
We distinguish (A) tasks directly related to recommendation, (B) tasks that are useful \emph{for} recommendation (training/evaluation/representation), and (C) adjacent non-RecSys agent tasks that sometimes appear in ARS pipelines.
Throughout the survey we treat (A) as primary, include (B) when it is used to improve recommendation, and discuss (C) only when it is tightly coupled to recommendation objectives or evaluation.

\subsection{Positioning with related surveys}
\label{subsec:related-surveys}

A fast-growing body of surveys has begun to chart LLMs, foundation models, and agents for recommendation, but with different emphases.
Broadly, existing works fall into two groups: (i) surveys on LLM-/FM-enhanced recommenders and (ii) surveys that explicitly foreground LLM agents or agentic paradigms.
Our survey is positioned at their intersection but adopts a distinct autonomy-centric, RS-specific viewpoint. This survey adopts an explicitly \textbf{autonomy-centered, RS-specific} perspective and seeks to unify agentic recommender systems across roles and implementation styles. The comparision with related surveys are summarized in Table~\ref{tab:survey-comparison}. 

\begin{table}[!t]
\small
  \caption{Positioning our survey with respect to existing surveys on LLMs, foundation models, and agentic recommender systems.}
  \label{tab:survey-comparison}
  \centering
  \begin{tabularx}{\linewidth}{@{}p{3.0cm}p{2.5cm}p{3.5cm}X@{}}
    \toprule
    \textbf{Survey} & \textbf{Primary scope} & \textbf{Main perspective} & \textbf{Gap w.r.t.\ our agentic LoA view} \\
    \midrule
    \citet{wu2024survey} (2024) &
    LLM-based RS &
    Discriminative vs.\ generative LLMs; pre-training / fine-tuning / prompting taxonomies &
    Focuses on LLM models; no explicit autonomy levels or multi-agent orchestration framing. \\
    \addlinespace[0.3em]
    \citet{zhao2024recommender} (2024) &
    LLM-enhanced RS &
    End-to-end view across data, model, and application layers &
    Does not differentiate retrieval-augmented, tool-using, and planning agents along an autonomy axis. \\
    \addlinespace[0.3em]
    \citet{li2024large} (2024) &
    Generative recommendation &
    LLMs as generators that directly output recommended items &
    Agent memory, tool use, and multi-agent collaboration are largely out of scope. \\
    \addlinespace[0.3em]
    \citet{huang2025survey} (2025a) &
    FM-powered RS &
    Feature-based vs.\ generative vs.\ agentic FM integration paradigms &
    ``Agentic RS'' is one paradigm; no level-wise mapping to concrete autonomy/design patterns. \\
    \addlinespace[0.3em]
    \citet{zhang2025survey} (2025) &
    Agents for RS \& search &
    Role-based taxonomy of LLM agents in IR (interface, optimizer, simulator, etc.) &
    Jointly covers search and RS; lacks an RS-tailored autonomy framework and level-wise mapping. \\
    \addlinespace[0.3em]
    \citet{peng2025survey} (2025) &
    LLM agents for RS &
    Scenario-centric paradigms: recommender-, interaction-, and simulation-oriented agents &
    Autonomy levels and planning workflows are mostly implicit. \\
    \addlinespace[0.3em]
    \citet{huang2025towards} (2025b) &
    Agentic RS perspective &
    Four-level evolution from static RS to agentic RS; multimodal LLMs and open challenges &
    Forward-looking; does not systematically catalogue existing ARS across fine-grained autonomy levels and tasks. \\
    \addlinespace[0.3em]
    \citet{zhu2025recommender} (2024/2025) &
    RS \& LLM agents (two-way) &
    RS for agents and agents for RS; strong focus on trustworthiness &
    Component- and trustworthiness-centric; lacks a unified LoA mapping of concrete RS architectures and roles. \\
    \addlinespace[0.3em]
    \citet{maragheh2025future} (2025) &
    Multi-agent RS (broad) &
    Definitions, coordination patterns, and system-level open challenges &
    Not LLM-specific; complements our RS-specific LoA and role-oriented mapping of existing LLM-driven ARS. \\
    \bottomrule
  \end{tabularx}
\end{table}

\subsection{Search Methodology}

The agentic RS field is evolving rapidly; to ensure comprehensive coverage
we followed a systematic search strategy.  We queried major academic
databases and digital libraries, including DBLP, the ACM Digital
Library, IEEE~Xplore and arXiv, from 2018 through March 2026.  Search
keywords combined terms from recommender systems and agentic AI, such as \textcolor{black}{
        "agent recommend",
        "RAG recommend",
        "retrieval-augmented recommend",
        "agent personalization"
        and "recommend simulator"}. 
Because the technology is new, few
papers prior to 2022 discussed LLM agents; nevertheless we included
earlier works on conversational RS and multi--agent recommender systems
as precursors. 
Concretely, we collect the papers in the following three steps: 
1) \textbf{Inclusion criteria.} We included papers involving autonomous or semi-autonomous agents for recommendation, user simulation, or evaluation, while excluding works using foundation models only as feature extractors. 
2) \textbf{Iterative snowballing.} Starting from a seed set, we expanded the corpus through backward and forward citation tracking until no major new works were found. 
3) \textbf{Annotation and coding.} Each paper was annotated by agent role, architecture, autonomy level, modality, evaluation, and trustworthiness aspects.

\begin{figure}[t]
\setlength{\abovecaptionskip}{-0.0cm}
\setlength{\belowcaptionskip}{-0.0cm}
\centering
\includegraphics[
width=0.99\linewidth,
trim={0 0.62cm 0 0},
clip
]{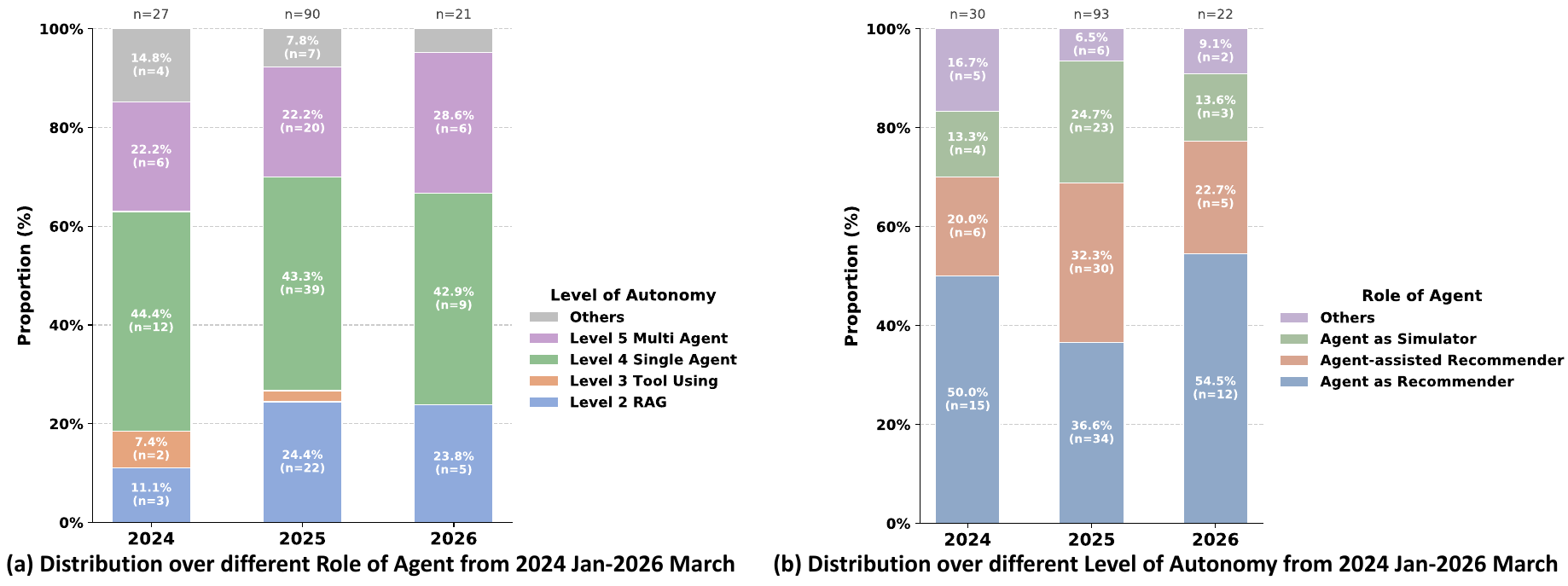}
\caption{Statistics of the reviewed literature from 2024 to March 2026. Left: distribution over different levels of autonomy. Right: distribution over different roles of agents. }
\label{fig:paper-statistics}
\end{figure}

\vspace{3pt}
\noindent\textbf{Literature statistics.} 
Based on the annotated results, we visualize the distributions of different agent roles and autonomy levels in Figure~\ref{fig:paper-statistics}, from which we have the following observations: 
(1) \textbf{Rapid growth in agentic recommendation}. The number of papers increased dramatically from approximately 27–30 in 2024 to 90–93 in 2025 (a roughly 3× growth), indicating that the agentic recommendation field has been attracting significant research interests.
(2) \textbf{Shift towards goal-driven automatic agentic recommendation.} Agent as Recommender is the dominant paradigm and continues to grow, while Agent as Simulator shows a clear upward trend, rising from 13.3\% in 2024 to 24.7\% in 2025. 
(3) \textbf{Shift towards higher-autonomy-level recommendation systems.} 
Level 4 (Single Agent) consistently dominates with over 40\%, while Level 5 (Multi-Agent) shows a steady upward trend (from 22.2\% in 2024 to 28.6\% in 2026).

\definecolor{myblue}{HTML}{A8DADC}
\begin{tcolorbox}[
    colback=myblue!20,        
    colframe=myblue!80!black, 
    boxrule=0.5pt,
    arc=2pt,
    left=6pt,
    right=6pt,
    top=6pt,
    bottom=6pt
]

\textbf{Core contributions of this survey}. 
Building on the above view, our main contributions are:
\begin{enumerate}[leftmargin=*]
    \item \textit{\textbf{An autonomy-centric definition and scope for agentic recommender systems.}}
    We formalize ARS as recommenders that embed LLM-powered agents with memory, tool use, and planning into the recommendation loop, clarifying boundaries with non-agentic LLM-based RS and earlier traditional RS.

    \item \textit{\textbf{A unified taxonomy (agent role $\times$ autonomy level).}}
    We use this taxonomy to systematically organize papers from the year of 2023 to early 2026 and reveal clusters and gaps across autonomy levels. For each level, we summarize the existing literature by identifying the key components and the high-level capabilities for autonomy across different tasks and scenarios. We also highlight the comparisons between each paradigm (\ie agent-assisted recommender, agent as recommender, and agent as simulator), and point out the key challenges with open research problems.

    \item \textit{\textbf{A detailed treatment of evaluation and benchmarking for ARS.}}
    We review offline and online metrics, LLM-as-judge protocols, and simulation-based assessments, and motivate agent-specific metrics for planning, tool use, efficiency, and trustworthiness.

\end{enumerate}

\end{tcolorbox}

\noindent\textbf{Organization.}
\S~\ref{sec:foundations-taxonomy} introduces foundational definitions (agents, tools, memory, workflows) and the taxonomies. 
Sections~\ref{sec:agent-assisted}-\ref{sec:agent_as_recommender} review agent-assisted recommenders, agent-as-recommenders, and agentic simulators (\S~\ref{sec:simulation})
Lastly, \S~\ref{sec:evaluation} focuses on evaluation and benchmarking, open challenges and future directions. 

\definecolor{softblue}{RGB}{220,230,242}
\definecolor{softgreen}{RGB}{226,239,218}
\definecolor{softpurple}{RGB}{229,224,236}
\definecolor{softyellow}{RGB}{255,242,204}
\definecolor{softred}{RGB}{242,220,219}
\definecolor{softgray}{RGB}{240,240,240}
\definecolor{softgold}{RGB}{235,190,115}
\definecolor{myyellow}{RGB}{255,242,204}

\tikzstyle{leaf}=[draw=black,
    rounded corners,minimum height=1em,
    text width=22.50em,
    text opacity=1,
    align=left,
    fill opacity=.3, text=black,font=\scriptsize,
    inner xsep=5pt, inner ysep=3pt,
    ]
\tikzstyle{leaf1}=[draw=black,
    rounded corners,minimum height=1em,
    text width=6.0em,
    text opacity=1, align=center,
    fill opacity=.5, text=black,font=\scriptsize,
    inner xsep=3pt, inner ysep=3pt,
    ]
\tikzstyle{leaf2}=[draw=black,
    rounded corners,minimum height=1em,
    text width=4.5em,
    text opacity=1, align=center,
    fill opacity=.8, text=black,font=\scriptsize,
    inner xsep=3pt, inner ysep=3pt,
    ]
\tikzstyle{leaf3}=[draw=black,
    rounded corners,minimum height=1em,
    text width=3.5em,
    text opacity=1, align=center,
    fill opacity=1.0, text=black,font=\scriptsize,
    inner xsep=3pt, inner ysep=3pt,
]

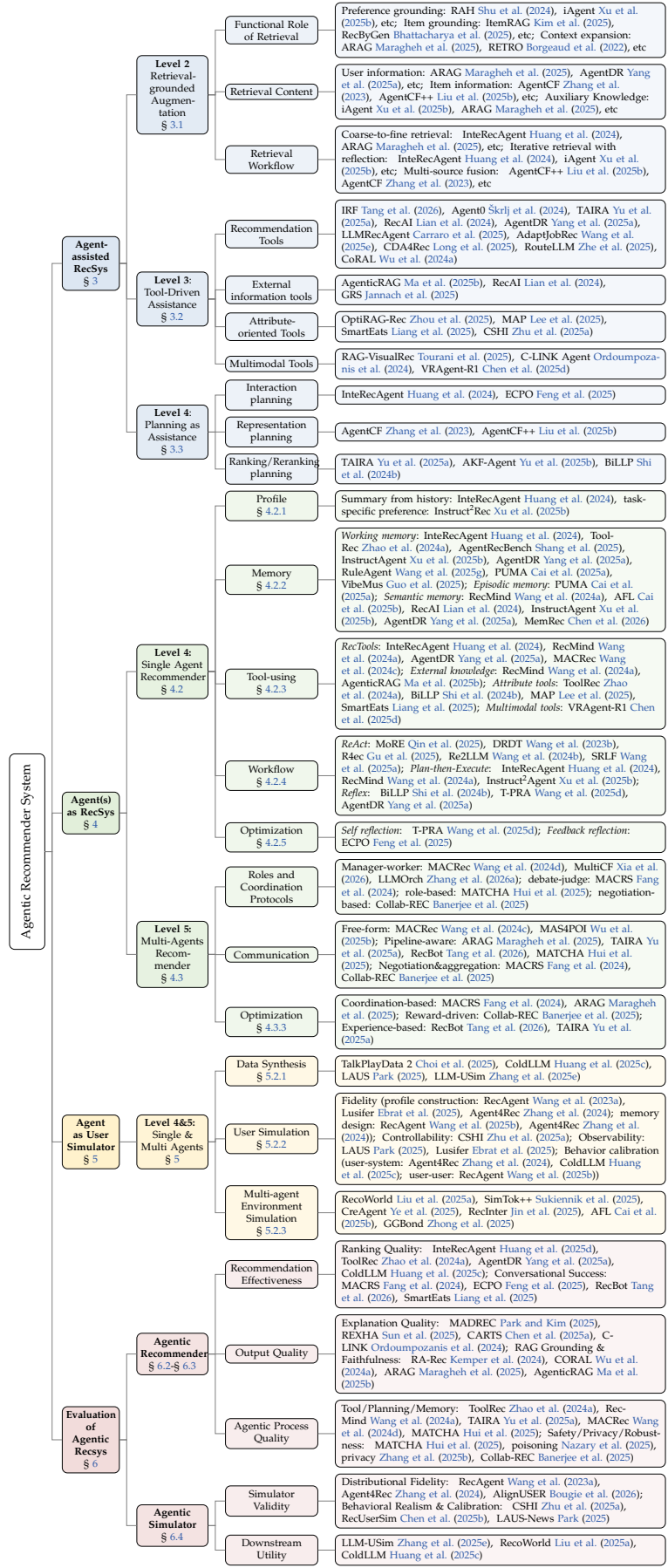
\begin{figure*}
\centering
\begin{adjustbox}{max totalsize={\textwidth}{\textheight},center}
\begin{forest}
  for tree={
    forked edges,
    grow=east,
    reversed=true,
    anchor=center,
    parent anchor=east,
    child anchor=west,
    base=center,
    font=\small,
    rectangle,
    draw=black,
    edge={black!50},
    rounded corners,
    minimum width=2em,
    minimum height=2.5em,
    s sep=5pt,
    inner xsep=3pt,
    inner ysep=1pt
  },
  where level=1{text width=5.5em}{},
  where level=2{text width=6em,font=\scriptsize}{},
  where level=3{font=\scriptsize}{},
  where level=4{font=\scriptsize}{},
  where level=5{font=\scriptsize}{},
  [Agentic Recommender System,rotate=90,anchor=north,inner xsep=8pt,inner ysep=3pt,edge={black!50},draw=black
    [\textbf{Agent-assisted} \\ \textbf{RecSys} \\ \S~\ref{sec:agent-assisted}, edge={black!50}, leaf3, fill=softblue,
      [\mbox{\textbf{Level 2}} \\ Retrieval-grounded Augmentation \\ \S~\ref{sec:assist-rag}, leaf2, fill=softblue,
        [Functional Role of Retrieval, leaf1, fill=softblue,
          [Preference grounding{: }RAH~\cite{shu2024rah}{, }
          iAgent~\cite{xu2025iagent}{, }etc{; }
          Item grounding{: }ItemRAG~\cite{kim2025itemrag}{, }
          RecByGen~\cite{bhattacharya2025recbygen}{, }etc{; }
          Context expansion{: }ARAG~\cite{maragheh2025arag}{, }
          RETRO~\cite{borgeaud2022improving}{, }etc
            ,leaf,fill=softblue]
         ],
        [Retrieval Content, leaf1, fill=softblue,
          [User information{: }ARAG~\cite{maragheh2025arag}{, }
          AgentDR~\cite{yang2025agentdr}{, }etc{; }
          Item information{: }AgentCF~\cite{zhang2023agentcf}{, }
          AgentCF++~\cite{liu2025agentcf++}{, }etc{; }
          Auxiliary Knowledge{: }iAgent~\cite{xu2025iagent}{, }
          ARAG~\cite{maragheh2025arag}{, }etc
            ,leaf,fill=softblue]
         ]
         [Retrieval Workflow, leaf1, fill=softblue,
            [Coarse-to-fine retrieval{: }
             InteRecAgent~\cite{huang2024interecagent}{, }
             ARAG~\cite{maragheh2025arag}{, }etc{; }
             Iterative retrieval with reflection{: }
             InteRecAgent~\cite{huang2024interecagent}{, }
             iAgent~\cite{xu2025iagent}{, }etc{; }
             Multi-source fusion{: }
             AgentCF++~\cite{liu2025agentcf++}{, }
             AgentCF~\cite{zhang2023agentcf}{, }etc
            ,leaf,fill=softblue]
         ]
      ]
      [\textbf{Level 3}:\\ Tool-Driven Assistance \\ \S~\ref{sec:assist-tools}, leaf2, fill=softblue,
        [Recommendation Tools, leaf1, fill=softblue,
          [IRF~\cite{tang2025interactive}{, }
          Agent0~\cite{vskrlj2024agent0}{, }
          TAIRA~\cite{yu2025thought}{, }
          RecAI~\cite{lian2024recai}{, }
          AgentDR~\cite{yang2025agentdr}{, }
          LLMRecAgent~\cite{carraro2025large}{, }
          AdaptJobRec~\cite{wang2025adaptjobrec}{, }
          CDA4Rec~\cite{long2025cloud}{, }
          RouteLLM~\cite{zhe2025constraint}{, }
          CoRAL~\cite{wu2024coral}
            ,leaf,fill=softblue]
         ],
        [External information tools, leaf1, fill=softblue,
          [AgenticRAG~\cite{ma2025agenticrag}{, }
          RecAI~\cite{lian2024recai}{, }
          GRS~\cite{jannach2025rethinking}
            ,leaf,fill=softblue]
         ]
         [Attribute-oriented Tools, leaf1, fill=softblue,
          [OptiRAG-Rec~\cite{zhou2025efficiency}{, }
          MAP~\cite{lee2025map}{, }
          SmartEats~\cite{liang2025smarteats}{, }
          CSHI~\cite{zhu2025llm}
            ,leaf,fill=softblue]
         ]
         [Multimodal Tools, leaf1, fill=softblue,
          [RAG-VisualRec~\cite{tourani2025rag}{, }
          C-LINK Agent~\cite{ordoumpozanis2024c}{, }
          VRAgent-R1~\cite{chen2025vragent}
            ,leaf,fill=softblue]
         ]
      ]
      [\textbf{Level 4}:\\ Planning as Assistance \\ \S~\ref{sec:assist-planning}, leaf2, fill=softblue,
        [Interaction planning, leaf1, fill=softblue,
          [InteRecAgent~\cite{huang2024interecagent}{, }
          ECPO~\cite{feng2025expectation}
            ,leaf,fill=softblue]
         ]
         [Representation planning, leaf1, fill=softblue,
          [AgentCF~\cite{zhang2023agentcf}{, }
          AgentCF++~\cite{liu2025agentcf++}
            ,leaf,fill=softblue]
         ]
         [Ranking/Reranking planning, leaf1, fill=softblue,
          [TAIRA~\cite{yu2025thought}{, }
          AKF-Agent~\cite{yu2025intelligent_agent}{, }
          BiLLP~\cite{shi2024planner}
            ,leaf,fill=softblue]
         ]
      ]
    ]
    [\textbf{Agent(s) as RecSys} \\ \S~\ref{sec:agent_as_recommender}, edge={black!50}, leaf3, fill=softgreen,
      [\mbox{\textbf{Level 4:}} \\ Single Agent \\ Recommender \\ \S~\ref{subsec:agent_rec_components}, leaf2, fill=softgreen,
       [Profile \\ \S~\ref{subsubsec:agent_rec_profile}, leaf1, fill=softgreen,
          [Summary from history{: }InteRecAgent~\cite{huang2024interecagent}{, }
          task-specific preference{: }Instruct$^2$Rec~\cite{xu2025iagent}
          ,leaf,fill=softgreen]
         ],
        [Memory \\ \S~\ref{subsubsec:agent_rec_memory}, leaf1, fill=softgreen,
          [\textit{Working memory}{: }InteRecAgent~\cite{huang2024interecagent}{, }
          ToolRec~\cite{zhao2024toolrec}{, }
          AgentRecBench~\cite{shang2025agentrecbench}{, }
          InstructAgent~\cite{xu2025iagent}{, }
          AgentDR~\cite{yang2025agentdr}{, }
          RuleAgent~\cite{wang2025ruleagent}{, }
          PUMA~\cite{cai2025large}{, }
          VibeMus~\cite{guo2025vibemus}{; }
          \textit{Episodic memory}{: }PUMA~\cite{cai2025large}{; }
          \textit{Semantic memory}{: }RecMind~\cite{wang2024recmind}{, }
          AFL~\cite{cai2025agentic}{, }
          RecAI~\cite{lian2024recai}{, }
          InstructAgent~\cite{xu2025iagent}{, }
          AgentDR~\cite{yang2025agentdr}{, }
          MemRec~\cite{chen2026memrec}
          ,leaf,fill=softgreen]
         ]
         [Tool-using \\ \S~\ref{subsubsec:agent_rec_tools}, leaf1, fill=softgreen,
          [\textit{RecTools}{: }InteRecAgent~\cite{huang2024interecagent}{, }
          RecMind~\cite{wang2024recmind}{, }
          AgentDR~\cite{yang2025agentdr}{, }
          MACRec~\cite{macrec}{; }
          \textit{External knowledge}{: }RecMind~\cite{wang2024recmind}{, }
          AgenticRAG~\cite{ma2025agenticrag}{; }
          \textit{Attribute tools}{: }ToolRec~\cite{zhao2024toolrec}{, }
          BiLLP~\cite{shi2024planner}{, }
          MAP~\cite{lee2025map}{, }
          SmartEats~\cite{liang2025smarteats}{; }
          \textit{Multimodal tools}{: }VRAgent-R1~\cite{chen2025vragent}
          ,leaf,fill=softgreen]
         ]
         [Workflow \\ \S~\ref{subsubsec:agent_rec_workflow}, leaf1, fill=softgreen,
          [\textit{ReAct}{: }MoRE~\cite{more}{, }
          DRDT~\cite{drdt}{, }
          R4ec~\cite{gu2025r}{, }
          Re2LLM~\cite{re2llm}{, }
          SRLF~\cite{srlf}{; }
          \textit{Plan-then-Execute}{: }
          InteRecAgent~\cite{huang2024interecagent}{, }
          RecMind~\cite{wang2024recmind}{, }
          Instruct$^2$Agent~\cite{xu2025iagent}{; }
          \textit{Reflex}{: }
          BiLLP~\cite{shi2024planner}{, }
          T-PRA~\cite{wang2025t-pra}{, }
          AgentDR~\cite{yang2025agentdr}
          ,leaf,fill=softgreen]
         ]
         [Optimization \\ \S~\ref{subsubsec:agent_rec_optimization}, leaf1, fill=softgreen,
          [\textit{Self reflection}{: }
          T-PRA~\cite{wang2025t-pra}{; }
          \textit{Feedback reflection}{: }
          ECPO~\cite{feng2025expectation}
          ,leaf,fill=softgreen]
         ]
      ]
      [\mbox{\textbf{Level 5:}} \\ Multi-Agents Recommender \\ \S~\ref{subsec:agent_rec_multi}, leaf2, fill=softgreen,
       [Roles and Coordination Protocols, leaf1, fill=softgreen,
          [Manager-worker{: }MACRec~\cite{wang2024macrec}{, }
          MultiCF~\cite{xia2026multi}{, }
          LLMOrch~\cite{zhang2026llms}{; }
          debate-judge{: }MACRS~\cite{fang2024multi}{; }
          role-based{: }MATCHA~\cite{hui2025matcha}{; }
          negotiation-based{: }Collab-REC~\cite{banerjee2025collab}
            ,leaf,fill=softgreen]
         ],
        [Communication, leaf1, fill=softgreen,
          [Free-form{: }MACRec~\cite{macrec}{, }
          MAS4POI~\cite{wu2025mas4poi}{; }
          Pipeline-aware{: }ARAG~\cite{maragheh2025arag}{, }
          TAIRA~\cite{yu2025thought}{, }
          RecBot~\cite{tang2025interactive}{, }
          MATCHA~\cite{hui2025matcha}{; }
          Negotiation\&aggregation{: }MACRS~\cite{fang2024multi}{, }
          Collab-REC~\cite{banerjee2025collab}
            ,leaf,fill=softgreen]
         ]
         [Optimization \\ \S~\ref{subsubsec:multi-agent-optimization}, leaf1, fill=softgreen,
          [Coordination-based{: }MACRS~\cite{fang2024multi}{, }
          ARAG~\cite{maragheh2025arag}{; }
          Reward-driven{: }Collab-REC~\cite{banerjee2025collab}{; }
          Experience-based{: }RecBot~\cite{tang2025interactive}{, }
          TAIRA~\cite{yu2025thought}
            ,leaf,fill=softgreen]
         ]
      ]
    ]
    [\textbf{Agent as User Simulator} \\ \S~\ref{sec:simulation}, edge={black!50}, leaf3, fill=myyellow,
      [\mbox{\textbf{Level 4\&5: }} \\ Single \& Multi Agents \\ \S~\ref{sec:simulation}, leaf2, fill=myyellow,
        [Data Synthesis \\ \S~\ref{subsubsec:simulation-data-synthesis}, leaf1, fill=myyellow,
          [TalkPlayData~2~\cite{choi2025talkplaydata}{, }
          ColdLLM~\cite{huang2025large}{, }
          LAUS~\cite{park2025llm}{, }
          LLM-USim~\cite{zhang2025llm}
            ,leaf,fill=myyellow]
         ]
        [User Simulation \\ \S~\ref{subsubsec:simulation-single-agent-simulator}, leaf1, fill=myyellow,
          [Fidelity (profile construction{: }RecAgent~\cite{wang2023recagent}{, }
          Lusifer~\cite{ebrat2025lusifer}{, }
          Agent4Rec~\cite{zhang2024agent4rec}{; }
          memory design{: }RecAgent~\cite{recagent}{, }
          Agent4Rec~\cite{zhang2024agent4rec}){; }
          Controllability{: }CSHI~\cite{zhu2025llm}{; }
          Observability{: }LAUS~\cite{park2025llm}{, }
          Lusifer~\cite{ebrat2025lusifer}{; }
          Behavior calibration (user-system{: }Agent4Rec~\cite{zhang2024agent4rec}{, }
          ColdLLM~\cite{huang2025large}{; }
          user-user{: }RecAgent~\cite{recagent})
            ,leaf,fill=myyellow]
         ]
         [Multi-agent \\ Environment \\ Simulation \\ \S~\ref{subsubsec:simulation-closed-loop-multi-agent-environment}, leaf1, fill=myyellow,
          [RecoWorld~\cite{liu2025recoworld}{, }
          SimTok++~\cite{sukiennik2025simulating}{, }
          CreAgent~\cite{ye2025creagent}{, }
          RecInter~\cite{jin2025recinter}{, }
          AFL~\cite{cai2025agentic}{, }
          GGBond~\cite{zhong2025ggbond}
            ,leaf,fill=myyellow]
         ]
      ]
    ]
    [\textbf{Evaluation of Agentic Recsys} \\ \S~\ref{sec:evaluation}, edge={black!50}, leaf3, fill=softred,
      [{\textbf{Agentic \\ Recommender}} \\ \S~\ref{subsec:outcome-eval}-\S~\ref{subsec:rag-eval}, leaf2, fill=softred,
        [Recommendation \\ Effectiveness , leaf1, fill=softred,
          [Ranking Quality{: }
          InteRecAgent~\cite{huang2025recommender}{, }
          ToolRec~\cite{zhao2024toolrec}{, }
          AgentDR~\cite{yang2025agentdr}{, }
          ColdLLM~\cite{huang2025large}{; }
          Conversational Success{: }
          MACRS~\cite{fang2024multi}{, }
          ECPO~\cite{feng2025expectation}{, }
          RecBot~\cite{tang2025interactive}{, }
          SmartEats~\cite{liang2025smarteats}
            ,leaf,fill=softred]
         ],
        [Output Quality, leaf1, fill=softred,
          [Explanation Quality{: }
          MADREC~\cite{park2025madrec}{, }
          REXHA~\cite{sun2025retrieval}{, }
          CARTS~\cite{chen2025carts}{, }
          C-LINK~\cite{ordoumpozanis2024c}{; }
          RAG Grounding \& Faithfulness{: }
          RA-Rec~\cite{kemper2024retrieval}{, }
          CORAL~\cite{wu2024coral}{, }
          ARAG~\cite{maragheh2025arag}{, }
          AgenticRAG~\cite{ma2025agenticrag}
            ,leaf,fill=softred]
         ],
        [Agentic Process \\ Quality, leaf1, fill=softred,
          [Tool{/}Planning{/}Memory{: }
          ToolRec~\cite{zhao2024toolrec}{, }
          RecMind~\cite{wang2024recmind}{, }
          TAIRA~\cite{yu2025thought}{, }
          MACRec~\cite{wang2024macrec}{, }
          MATCHA~\cite{hui2025matcha}{; }
          Safety{/}Privacy{/}Robustness{: }
          MATCHA~\cite{hui2025matcha}{, }
          poisoning~\cite{nazary2025stealthy}{, }
          privacy~\cite{zhang2025towards}{, }
          Collab-REC~\cite{banerjee2025collab}
            ,leaf,fill=softred]
         ]
      ]
      [{\textbf{Agentic \\ Simulator}} \\ \S~\ref{subsec:simulator-eval}, leaf2, fill=softred,
        [Simulator \\ Validity, leaf1, fill=softred,
          [Distributional Fidelity{: }
          RecAgent~\cite{wang2023recagent}{, }
          Agent4Rec~\cite{zhang2024agent4rec}{, }
          AlignUSER~\cite{bougie2026alignuser}{; }
          Behavioral Realism \& Calibration{: }
          CSHI~\cite{zhu2025llm}{, }
          RecUserSim~\cite{chen2025recusersim}{, }
          LAUS-News~\cite{park2025llm}
            ,leaf,fill=softred]
         ],
        [Downstream \\ Utility, leaf1, fill=softred,
          [LLM-USim~\cite{zhang2025llm}{, }
          RecoWorld~\cite{liu2025recoworld}{, }
          ColdLLM~\cite{huang2025large}
            ,leaf,fill=softred]
         ]
      ]
    ]
  ]
\end{forest}
\end{adjustbox}
\caption{Overview structure of this survey.}
\label{fig:tree}
\end{figure*}

\section{Foundations and Taxonomy of Agentic Recommender Systems}
\label{sec:foundations-taxonomy}

{This section unifies two goals: (i) formalize agentic recommendation as an \emph{interactive decision process} and
define the entities we study (agents and agentic recommender systems) in \S \ref{subsec:recommender_as_interactive_decision}, and (ii) introduce a taxonomy that can organize
a rapidly growing literature beyond conventional RS task/method categorizations in \S \ref{subsec:two-axis-taxonomy-rewrite}.}

\subsection{Recommender systems as interactive decision processes}
\label{subsec:recommender_as_interactive_decision}

Classical recommender systems are often presented as static mapping functions that produce a ranked list given a user and context. Agentic recommender systems (ARS), in contrast, are more naturally modeled as \textit{interactive decision processes} in which the system repeatedly observes a user’s requests and feedback, consults internal state and external resources, and decides what to do next. This shift is not merely a different UI; it changes what should be considered “the recommender”: the recommender becomes an \textit{entity that chooses actions} (\eg ask a question, retrieve candidates, apply constraints, call tools, verify a plan, generate an explanation), rather than a single-shot ranker.

To make this explicit, we view recommendation as proceeding in discrete turns \(t = 1, 2, \dots\). Let \(u_t\) denote the user input at turn \(t\) (which could be a query, constraints, or feedback). Let \(h_t = (u_1, y_1, \dots, u_t)\) denote the interaction history up to the current turn, where \(y_t\) is the system output (\eg recommendation response, clarification question, or tool-triggered result). In agentic settings, the system also maintains an internal state \(s_t\), which includes {user/item profiles}, {memory}, and other latent variables required for long-horizon consistency. 
The system chooses an action \(a_t\) from an action space \(\mathcal{A}\). Importantly, \(\mathcal{A}\) is not restricted to “recommend item \(i\).” 
It typically includes: 
\textit{dialogue actions} (ask, confirm, explain, negotiate trade-offs), \textit{tool actions} (retrieve, rank, filter, query a KB, browse the web), \textit{environment actions} (\eg add to cart, reserve, schedule) when in scope.

In the following, we define the LLM agent and agentic recommender, respectively. 
\vspace{2pt}
\begin{definition}[LLM Agent for Recommendation]
An LLM agent is an entity that repeatedly (i) \textbf{observes} an input comprising user signals and environment/tool outputs, (ii) \textbf{reasons} to interpret goals and constraints, (iii) optionally \textbf{plans} a sequence of actions, (iv) \textbf{acts} by producing user-facing outputs and/or invoking tools, and (v) \textbf{updates} its state (profile/memory) based on new evidence.
\end{definition} 
We represent an LLM agent as $A = (\mathcal{L}_\theta, \pi, \mathcal{P}, \mathcal{M}, \mathcal{T})$, where $\mathcal{L}_\theta$ is the LLM (or foundation model) used as the main reasoning engine.
\(\pi\) is the {controller/policy} that determines how the LLM is prompted, how tool calls are formatted/parsed, and how multi-step execution proceeds (\eg ReAct). 
Different workflows are detailed in \S~\ref{subsubsec:agent_rec_workflow}. 
\(\mathcal{P}\) is the {profile module} (persistent traits and summaries of the user and/or items). 
\(\mathcal{M}\) is the {memory module} (a store of episodic and/or semantic information used across turns and sessions). 
\(\mathcal{T}\) is the {tool set} (retrieval, ranking services, search engines, KB queries, etc.), along with a tool interface that maps agent outputs to executable calls.

\begin{definition}[Agentic Recommender System, ARS]
An agentic recommender system is an interactive recommendation system that contains one or more LLM agents that can plan and act using tools and memory to produce personalized outcomes. 
We formalize: $ARS = (\mathcal{U}, \mathcal{I}, \mathcal{E}, \mathcal{R}, \mathcal{A})$, 
where \(\mathcal{U}\) is the set of users, \(\mathcal{I}\) is the item universe, \(\mathcal{E}\) is the external environment (web, APIs, KBs, simulators), \(\mathcal{R}\) denotes classical RS components (optional; present in agent-assisted ARS), and \(\mathcal{A}=\{A_1, \dots, A_n\}\) is the set of agents.
\end{definition}

This definition is intentionally broad enough to cover three recurring paradigms in the literature: \textbf{agent-assisted recommendation}, \textbf{agent-as-recommender}, and \textbf{agent-as-user-simulator}. In this survey, we describe this role-based partition and ties it to LoA levels, which is detailed in \S~\ref{subsec:two-axis-taxonomy-rewrite}.

\subsection{Why agentic systems require a new taxonomy}
\label{subsec:new-taxonomy-rationale}
{
Conventional taxonomies in recommender systems have focused on algorithmic families (matrix factorization versus neural networks), task categories (retrieval, ranking, reranking) or data modalities.  While such classifications were adequate for static pipelines, they struggle to describe the emergent landscape of \emph{agentic recommenders}.  In these systems the recommender is no longer a single‑shot scorer but an entity that can observe, reason, plan and act over multiple turns.  Agents can occupy different roles within the recommendation loop—sometimes augmenting a classical model with retrieval or constraint reasoning, sometimes replacing it entirely, and sometimes simulating users or environments to generate feedback for training and evaluation.  Two systems trained on the same dataset may therefore differ radically in how they treat an agent: one may rely on an agent merely to translate natural language into search queries, while another may delegate all decision‑making to the agent.  These differences affect the action space, interaction protocol, and evaluation methodology; they cannot be captured by algorithmic labels alone.  A new taxonomy must therefore capture both \emph{where} the agent is positioned within the recommendation pipeline and \emph{what} the agent is allowed to do, providing a framework that unifies agentic augmentation, agentic replacement, and agentic simulations.
}

\subsection{A two‑axis taxonomy: roles and autonomy levels}
\label{subsec:two-axis-taxonomy-rewrite}
{
To organise the burgeoning literature, we propose a two‑dimensional taxonomy that cross‑classifies agentic recommenders by \textbf{macro role}—the agent’s placement in the recommendation loop—and by \textbf{autonomy level}—the breadth of actions the agent can perform and the capabilities the agent can achieve. This structure clarifies relationships among systems that otherwise share similar datasets or metrics.
}

\begin{figure}[t]
\setlength{\abovecaptionskip}{0.02cm}
\setlength{\belowcaptionskip}{-0cm}
\centering
\includegraphics[width=0.99\linewidth]{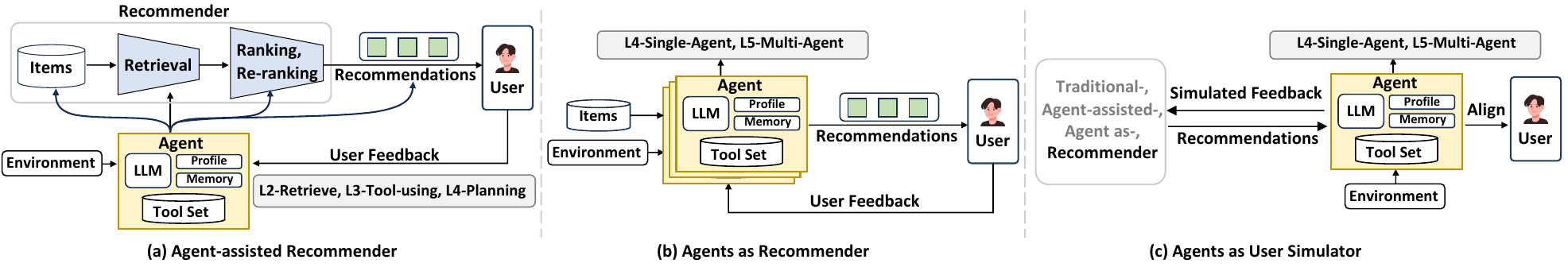}
\caption{Illustration of three paradigms of agentic recommender systems. {While all of these variations make use of tools, memory, and user profiles, the main difference between (a) and (b) lies in whether they involve a classical retrieval component—\eg dense retrieval or collaborative-style retrieval as in (a)—or whether this process is handled by a single agent/LLM without the help of auxiliary retrieval modules, as in (b). In (c), agents simulate user behavior and provide simulated feedback to analyze or improve the recommender system. Thus, for example, in this view, we observe an agentic RAG workflow, where the agent is assisted by external tools, an LLM, and memory components to support retrieval and generation within the recommender pipeline.}}
\label{fig:paradigm_comparison}
\end{figure}

\paragraph{\textbf{Macro roles.}} We distinguish three high‑level roles as illustrated in Figure~\ref{fig:paradigm_comparison}:
\begin{enumerate}
  \item \textbf{Agentic augmentation}.  The agent works alongside a classical recommender, improving specific stages such as preference elicitation, evidence retrieval, or result filtering.  The classical model still produces the final ranking.  Systems like \emph{RecMind} and \emph{iAgent} fall here, where the agent interprets natural‑language queries and invokes retrieval or ranking tools but does not itself decide the top items.
  \item \textbf{Agentic replacement}.  One or more agents assume full responsibility for recommendation.  They maintain user state, orchestrate tool calls, retrieve and rank items, and generate explanations.  Classical models, if present, appear as tools rather than the primary controller.  Examples include \emph{InteRecAgent}, \emph{MADREC} and \emph{DRDT}, where an LLM agent decides both what to recommend and how to justify it.
  \item \textbf{Agentic simulation}.  Agents model the behaviour of users or environments, generating clicks, ratings, critiques or dialogue acts for training, evaluation and robustness analysis.  Multi‑agent environments such as \emph{RecoWorld} and \emph{LAUS‑News} allow experiments with complex user–item dynamics without deploying on real users.
\end{enumerate}

\paragraph{\textbf{Autonomy levels.}} Orthogonal to role, autonomy measures how broadly an agent can act.  This survey focuses on:
\begin{enumerate}
  \item \textbf{L2 — Retrieval‑grounded assistance.}  The agent supplements a classical system by retrieving relevant evidence—user history, item attributes, domain knowledge—but does not orchestrate tools or plan.  This tier includes retrieval‑augmented generation frameworks such as ARAG that ground LLM outputs in evidence to reduce hallucinations.
  \item \textbf{L3 — Tool‑orchestrating assistance.}  The agent selects and invokes specialized tools implementing subroutines (retrievers, rankers, constraint solvers, search engines, multimodal handlers) and may choose among them.  ToolRec and iAgent exemplify this level.
  \item \textbf{L4 — Single‑agent planning.}  A single agent performs multi‑step reasoning: it decomposes tasks, calls tools sequentially, reflects on intermediate results and self‑corrects errors.  Controllers may follow \textsc{ReAct} loops, planner–executor frameworks or reflection mechanisms.  \emph{InteRecAgent} and \emph{Thought‑Augmented Planning} are characteristic examples.
  \item \textbf{L5 — Multi‑agent orchestration.}  Decision authority is distributed across specialised agents—planners, retrievers, rankers, critics, explainers and safety monitors—that communicate via explicit protocols.  Manager–worker, debate–judge and negotiation schemes allow agents to coordinate.  Systems such as \emph{MACRec}~\cite{macrec}, \emph{MACRS}~\cite{fang2024multi}, \emph{Collab‑REC}~\cite{banerjee2025collab}, \emph{RouteLLM}~\cite{zhe2025constraint} and \emph{MATCHA}~\cite{hui2025matcha} illustrate this tier.
\end{enumerate}

Crossing these axes reveals the diversity of agentic systems.  For instance, a retrieval‑augmented assistant that re‑ranks items from a classical model belongs to the “agentic augmentation” macro role with L3 autonomy, whereas a multi‑agent simulation environment that models users and producers belongs to the “agentic simulation” role with L5 autonomy.  This cross‑classification clarifies both how agents interact with classical components and how much autonomy they exercise.

\section{Agentic Augmentation: Agent-Assisted Recommender Systems}
\label{sec:agent-assisted}

{Agent-assisted recommender systems insert an LLM agent \emph{around} a pre-existing recommendation pipeline to \textit{augment} specific sub-tasks (\eg retrieval/grounding, intent understanding, explanation, or constraint enforcement), while the core recommender remains the primary decision module.
In the language of autonomy, this category largely subsumes \textbf{L2 (retrieval-augmented)} and \textbf{L3 (tool-using)} systems: the agent improves the pipeline without fully replacing it, and the output is typically still anchored to classic recommendation primitives such as candidate generation and ranking.
This “assistance” framing matters because it changes what we optimize and what we measure: rather than asking whether an agent can recommend end-to-end, we ask which \emph{interventions} improve user and system outcomes, and how to evaluate those interventions reliably.}

{\paragraph{What tasks do agents assist with?}
Across the recent literature, agent-assisted systems repeatedly target a small set of recurring tasks, which can be grouped by \emph{where} they intervene in the recommendation loop:}

\begin{itemize}
    \item {\textbf{Understanding and control}: interpreting natural-language intent and constraints, eliciting missing preferences, and translating user commands into structured control signals (\eg conversational steering, controllable constraints) \cite{tang2025interactive,feng2025expectation,huang2025recommender,chen2025multi,zeng2024automated,luo2025rallrec+}.} 
    \item {\textbf{Grounding and knowledge access}: retrieving relevant information to reduce hallucinations and increase specificity (user-history snippets, item semantics such as attributes/reviews, external knowledge such as domain facts), typically via RAG-style components} \cite{tang2025interactive,feng2025expectation,huang2025recommender}. 
    \item {\textbf{Candidate operations and reranking}: generating candidate sets, filtering candidates under constraints, and reranking by richer semantic criteria (\eg compatibility reasoning or cross-domain transfer), often by coupling agents to RecTools \cite{zhao2024toolrec,yang2025agentdr,liu2025agentcf++,alamri2025leveraging}.} 
    \item {\textbf{Explanation and interaction quality}: producing grounded explanations and multi-turn justifications; summarizing rationales; clarifying trade-offs; adapting explanations to context \cite{sun2025retrieval,park2025madrec}.}
    \item {\textbf{Governance}: enforcing safety/fairness constraints, inserting intermediate representations or shields, and providing “guardrails” that sit between user, agent, and recommender \cite{xu2025iagent}; leveraging agent as engineer to update the recommendation strategies~\cite{lao2026agentx，wang2026self}.}
\end{itemize}

{Tasks often span multiple stages. For example, a multi-turn conversational assistant may (i) interpret an instruction, (ii) retrieve relevant history, (iii) call a ranking tool, and (iv) generate a justification; tool use and retrieval become \emph{mechanisms} that enable the task rather than ends in themselves.}

\subsection{Retrieval-Augmented Assistance}
\label{sec:assist-rag}

\begin{figure}[t]
\setlength{\abovecaptionskip}{0.02cm}
\setlength{\belowcaptionskip}{-0.3cm}
\centering
\includegraphics[width=0.99\linewidth]{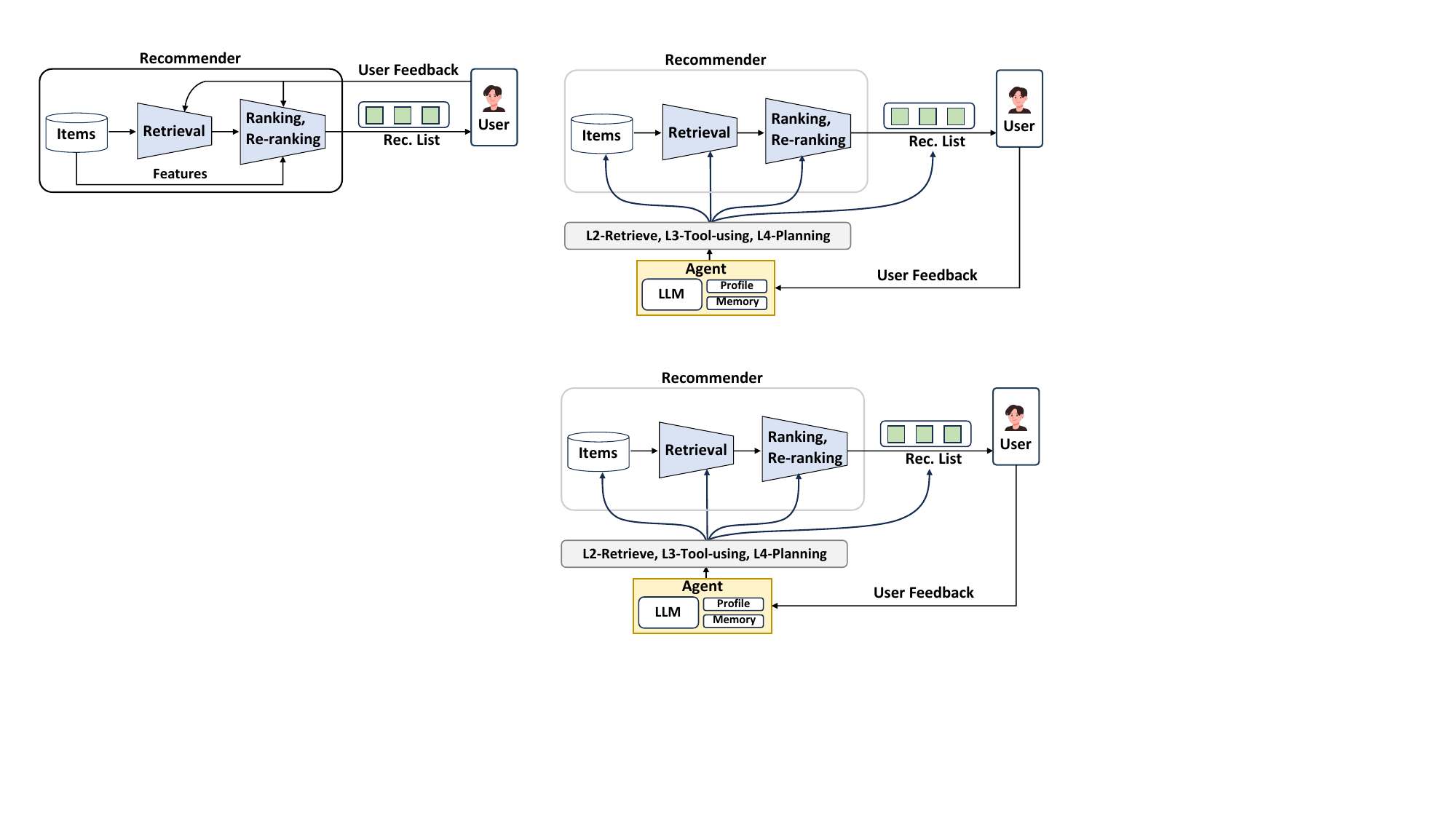}
\caption{Illustration of three aspects of Retrieval-Augmented Assistance for Agent-assisted Recommender Systems.}
\label{fig:agent_assist_rag_tool_planning}
\end{figure}

{Retrieval augmentation is the dominant pattern for L2 systems and a key building block for L3 systems. While RAG is broadly used in NLP, in recommender systems the retrieved evidence must align with recommendation structure: it may contain \emph{user-side} information (profiles, histories, constraints), \emph{item-side} information (descriptions, reviews, attributes), or \emph{auxiliary knowledge} (domain facts, knowledge graphs, taxonomies).
Following the original survey’s framing, it is useful to organize retrieval-augmented recommendation through three complementary lenses: 

{{(i) \textbf{\textit{Functional role}}: why retrieval is invoked.}
Common functional roles include:
(1) \textbf{Preference grounding} (retrieve history snippets to ground intent and reduce inconsistency across turns). 
For example, RAH~\cite{shu2024rah} enhances the retrieval stage by querying a structured personality library to extract user-specific preferences, goals, and value orientations. These retrieved signals are then used to filter and prioritize candidate items, enabling more accurate and personalized candidate generation before downstream ranking.
iAgent~\cite{xu2025iagent} retrieves external knowledge through tool invocation while simultaneously leveraging the internal knowledge of LLMs, thereby assisting the reranker in further refining the ordering of an already ranked item list.
(2) \textbf{Item grounding} (retrieve item evidence to support explanations and avoid fabrication) \cite{kim2025itemrag,bhattacharya2025recbygen},
and (3) \textbf{Context expansion} (retrieve domain knowledge for niche domains such as health, finance, or travel) \cite{maragheh2025arag,borgeaud2022improving,zhao2025webrec,cho2025marc,banerjee2024enhancing, rao2024ramo}.
This perspective clarifies what “success” means: \eg for preference grounding we care about stability and faithfulness to user history, whereas for context expansion we care about factuality and usefulness.} 

{{(ii) \textbf{\textit{Retrieved content}}: what is retrieved.}
Retrieved content typically falls into:
\textbf{user information} (profiles, histories, constraints)~\cite{huang2024interecagent,shu2024rah,wang2024recmind,yu2025intelligent_agent,maragheh2025arag,kim2026leveraging, wei2025enhanced, ao2025retrieval, ning2026retrieval, yang2025retrieval}. 
User information is typically divided into long-term and short-term preferences and represented as interaction histories or textual preference evidence. 
For example, ARAG~\cite{maragheh2025arag} collects textual cues from both the current session and long-term logs to summarize user intent and support re-ranking. AgentCF~\cite{zhang2023agentcf} retrieves user memory and item memory to enrich preference representations and item semantics. AgentDR~\cite{yang2025agentdr} further incorporates tool-applicability memory, improving the automation level of agent-assisted recommendation. 
\textbf{item semantics} (attributes/reviews; multimodal content where available)~\cite{zhang2023agentcf,liu2025agentcf++,kim2025itemrag, wang2025knowledge, qiu2025graph, kieu2025keyword, li2025g, xu2025rallrec, yang2025cold,balachandran2025visiorag,wei2025learning}; for example, AgentCF~\cite{zhang2023agentcf} retrieves item memory to enrich item-side semantics, while AgentCF++~\cite{liu2025agentcf++} further leverages cross-domain semantic evidence to improve candidate understanding. 
and \textbf{auxiliary knowledge} (external corpora, KGs, or curated domain sources)  \cite{maragheh2025arag,xu2025iagent,borgeaud2022improving,meng2025balancing}; for example, iAgent~\cite{xu2025iagent} queries outside knowledge sources to refine reranking. 
This choice affects privacy and leakage risk: user-side retrieval often contains sensitive personal data, while auxiliary knowledge retrieval raises provenance and licensing issues.}

{{(iii) \textbf{\textit{Retrieval flows}}: how retrieval is sequenced.}
Beyond “retrieve once then generate,” recent systems increasingly use multi-step flows, such as:
1) {coarse-to-fine retrieval} (retrieve broad candidates, then refine by constraints) \cite{huang2024interecagent,maragheh2025arag,li2025rallm},
2) {iterative retrieval with reflection} (retrieve, draft, detect missing evidence, retrieve again) \cite{huang2024interecagent,xu2025iagent,zhang2025sarrec},
or 3) {multi-source fusion} (combine user-history retrieval with item-attribute retrieval) \cite{zhang2023agentcf,liu2025agentcf++,zeng2024federated}.
These flows implicitly introduce \emph{planning} and should be evaluated at the trajectory level, not only by final accuracy.} 

{\paragraph{\textbf{Key limitations.}}
RAG-based assistance can still fail via (i) \emph{retrieval bias} (frequent or easy-to-retrieve patterns dominate), (ii) \emph{context overload} (critical evidence is “lost in the middle”), and (iii) \emph{spurious grounding} (retrieved text gives false confidence without causal relevance) \cite{maragheh2025arag,sun2025retrieval,xu2025iagent,liu2024lost,lichtenberg2024large,sun2025cirr,zhang2025customizedretrievalaugmentedgenerationllm, tandon2025evaluating, sayana2025beyond, lin2025strajrag, cheng2025education, meng2025kerag_r}.
These motivate evaluation protocols that explicitly test robustness to retrieval noise and long-context failure (Sec.~\ref{sec:evaluation}).  
Compared with conventional retrieval systems, retrieval-augmented recommender systems incorporate external behavioral traces and world knowledge into the recommendation process, thereby enhancing personalization and generalization. 
However, their autonomy remains limited: while they are able to integrate long-term context and historical memory to some extent, they still primarily operate reactively rather than proactively, rely on single-turn retrieval with limited interaction flexibility, and utilize fixed pipelines that are difficult to adapt. 
Thus, retrieval-augmented recommender systems can be regarded as a transitional stage bridging static retrieval models and autonomous recommendation agents capable of strategic reasoning, context-aware planning, and adaptive learning.

\begin{table*}[!t]
\centering
\caption{Unified Summary of Retrieval-Augmented Assistance, Tool-Driven Assistance, and Planning as Assistance.}
\label{tab:unified_summary}
\resizebox{\textwidth}{!}{%
\begin{tabular}{p{2.5cm} p{3cm} p{3.5cm} p{5cm} p{5cm}}
\toprule
\textbf{Category} & \textbf{Dimension} & \textbf{Sub-type} & \textbf{Representative Works} & \textbf{High-level Capabilities} \\
\midrule

\multirow{9}{=}{\textbf{Retrieval-Augmented Assistance}}

& \multirow{3}{=}{\textbf{Functional Role} \newline (Why retrieval is invoked)}
& Preference Grounding
& RAH~\cite{shu2024rah}, iAgent~\cite{xu2025iagent}
& \multirow{9}{=}{\textbf{Proactivity:} Reactive \newline
\textbf{Context Awareness:} Memory (long-/short-term) \newline
\textbf{Interaction:} Single-turn retrieval \newline
\textbf{Adaptivity:} Static pipeline} \\

& & Item Grounding
& AgentCF~\cite{zhang2023agentcf}, AgentCF++~\cite{liu2025agentcf++}
& \\

& & Context Expansion
& ARAG~\cite{maragheh2025arag}, iAgent~\cite{xu2025iagent}
& \\

\cmidrule{2-4}

& \multirow{3}{=}{\textbf{Retrieved Content} \newline (What is retrieved)}
& User Information
& InteRecAgent~\cite{huang2024interecagent}, RAH~\cite{shu2024rah}, RecMind~\cite{wang2024recmind}, ARAG~\cite{maragheh2025arag}, AgentCF~\cite{zhang2023agentcf}, AgentDR~\cite{yang2025agentdr}
& \\

& & Item Semantics
& AgentCF~\cite{zhang2023agentcf}, AgentCF++~\cite{liu2025agentcf++}
& \\

& & Auxiliary Knowledge
& ARAG~\cite{maragheh2025arag}, iAgent~\cite{xu2025iagent}
& \\

\cmidrule{2-4}

& \multirow{3}{=}{\textbf{Retrieval Flows} \newline (How retrieval is sequenced)}
& Coarse-to-Fine
& InteRecAgent~\cite{huang2024interecagent}, ARAG~\cite{maragheh2025arag}
& \\

& & Iterative with Reflection
& InteRecAgent~\cite{huang2024interecagent}, iAgent~\cite{xu2025iagent}
& \\

& & Multi-source Fusion
& AgentCF~\cite{zhang2023agentcf}, AgentCF++~\cite{liu2025agentcf++}
& \\

\midrule

\multirow{4}{=}{\textbf{Tool-Driven Assistance}}

& \multirow{4}{=}{\textbf{Tool Taxonomy} \newline (What tools are used)}
& RecTools
& InteRecAgent~\cite{huang2024interecagent}, ToolRec~\cite{zhao2024toolrec}
& \multirow{4}{=}{\textbf{Proactivity:} Proactive (tool invocation) \newline
\textbf{Context Awareness:} Internal \& external memory \newline
\textbf{Interaction:} Multi-turn \newline
\textbf{Adaptivity:} Adaptive strategy} \\

& & External Info Tools
& iAgent~\cite{xu2025iagent}, RecMind~\cite{wang2024recmind}
& \\

& & Attribute-oriented Tools
& ToolRec~\cite{zhao2024toolrec}
& \\

& & Multimodal Tools
& HuggingGPT~\cite{shen2023hugginggpt}, Toolformer~\cite{schick2023toolformer}
& \\[14pt]

\midrule

\multirow{8}{=}{\textbf{Planning as Assistance}}

& \multirow{2}{=}{\textbf{Interaction} \newline \textbf{Planning}}
& Reflective Plan Revision
& InteRecAgent~\cite{huang2024interecagent}
& \multirow{8}{=}{\textbf{Planning:} Goal decomposition, tool orchestration \newline
\textbf{Reasoning:} Multi-step reasoning, reflection \newline
\textbf{Adaptivity:} Dynamic memory \& plan updates \newline
\textbf{Context Awareness:} Tools, history, interactions} \\

& & Expectation-driven Steering
& ECPO~\cite{feng2025expectation}
& \\[4pt]

\cmidrule{2-4}

& \multirow{2}{=}{\textbf{Representation} \newline \textbf{Planning}}
& Semantic Enrichment
& AgentCF~\cite{zhang2023agentcf}
& \\

& & Cross-domain Transfer
& AgentCF++~\cite{liu2025agentcf++}
& \\[4pt]

\cmidrule{2-4}

& \multirow{2}{=}{\textbf{Ranking/Reranking} \newline \textbf{Planning}}
& Reasoning-driven Reranking
& Intelligent Agent~\cite{yu2025intelligent_agent}
& \\

& & Reflective Learning
& Re2LLM~\cite{re2llm}, MoRE~\cite{more}, SRLF~\cite{srlf}, DRDT~\cite{drdt}
& \\

\bottomrule
\end{tabular}
}
\end{table*} 

\subsection{Tool-Driven Assistance}
\label{sec:assist-tools}

{Tool use is the hallmark of L3 systems: the agent does not merely condition on retrieved text, but \emph{acts} via API calls, database queries, retrieval services, rerankers, and other components.
In recommendation, tool use spans different stages, such as retrieval, ranking, and reranking. 
}

\paragraph{\textbf{Tool taxonomies and design choices.}}
A practical taxonomy separates:
(i) \textbf{RecTools} directly provides candidate sets or ranking lists (\eg traditional retrievers, rankers, candidate filters models); 
(ii) \textbf{external information tools} retrieve external knowledge and contextual signals (\eg web/domain search, knowledge base query tools); 
(iii) \textbf{attribute-oriented tools} filter items based on structured attributes or specific constraints (\eg feature calculators, KG completion, constraint checkers); 
and (iv) \textbf{multimodal tools} are used to understand multimodal features, or generate multimodal content to supplement recommendation tasks (\eg vision/audio encoders, captioners) \cite{zhao2024toolrec,huang2024interecagent,yang2025agentdr,shen2023hugginggpt,schick2023toolformer,shi2025ocgagent}. 
These tools can be invoked separately or sequentially in different contexts. 
For example, InteRecAgent~\cite{huang2024interecagent} calls recommendation tools to support task planning and execution, progressively refining the candidate pool before returning it to the LLM, which then delivers the final response to the user. 
In re-ranking, iAgent~\cite{xu2025iagent} queries external knowledge through search tools and forwards the retrieved information to a re-ranker to refine an already-ranked list. 
AgentDR exemplifies a tool-centric view where the agent learns to uncover implicit item--item relations by invoking a pipeline of tool operations, enabling dynamic recommendation under changing contexts \cite{yang2025agentdr}. 
{Tool selection policies can be hand-coded, prompted, or learned; recent work on modular ``hooks'' attempts to decouple tool invocation logic from prompts, improving maintainability and safety auditing \cite{de2024language}.
In practice, tool-driven assistance often succeeds not because the LLM is a strong ranker, but because it orchestrates structured components with better inductive bias (retrieval, ranking, graph search).

\subsection{Planning as Assistance}
\label{sec:assist-planning}

{Planning arises naturally once agents operate over multi-turn conversations and multi-step tool sequences. Importantly, planning is not only about conversation; it can appear in:
(1) \textbf{interaction planning} (\eg what to respond next, how to improve the recommendations) can be injected to decide what to ask next to the user, or be leveraged to generate intermediate plans to give guidance signals to improve recommender models that can align better with the user needs~\cite{huang2024interecagent,feng2025expectation}. 
(2) \textbf{Rrepresentation planning} (\eg what and how to update user preference and item descriptions) allows LLMs to construct richer, updated, and explainable user and item descriptions~\cite{zhang2023agentcf,liu2025agentcf++, xie2026agentictagger}.
(3) \textbf{Ranking/reranking planning} usually considers what constraints to satisfy first, or how to trade-off objectives, thus leading to a more constraint-compatible result that aligns with user preference at the same time~\cite{yu2025thought,yu2025intelligent_agent,shi2024large}.}
(4) \textbf{Optimization planning} leverages LLM agents to automatically diagnose algorithm limitations and update the recommendation strategies/pipelines iteratively~\cite{lao2026agentx,wang2026self}.

Existing literature has explored different types of planning to augment recommender systems. 
For example, InteRecAgent~\cite{huang2024interecagent} introduces explicit planning into the recommendation pipeline. 
By reflecting on user interactions, tool invocation plans, and intermediate recommendation outcomes, the agent continuously revises and optimizes its internal plans, thereby enhancing the system’s understanding of user intent and enabling more controllable and self-improving interactions. 
To enhance user/item representations, AgentCF~\cite{zhang2023agentcf} leverages automated user interaction histories and textual feedback to enrich the semantic representations of both users and items. By incorporating global collaborative information, it produces more comprehensive and fine-grained embeddings that offer a stronger foundation for downstream recommendation. In the ranking task, the Intelligent Agent~\cite{yu2025intelligent_agent} applies predefined reasoning rules to the pre-ranked list, performing step-by-step inference to assess whether each candidate item truly aligns with the user’s explicit or implicit preferences. This reasoning-driven refinement yields a more accurate and user-aligned final ranking than traditional scoring-based approaches.

{\paragraph{\textbf{Comparison between tool-driven and planning assistance (L3 $\rightarrow$ L4).}}
Compared with tool-driven agents, the single-agent planner stage exhibits stronger autonomy: it can actively set goals, break them into steps, and continuously observe, reflect, and adjust during execution; 
Together, these capabilities enable richer context awareness, proactive decision-making, flexible interaction, and continuous adaptivity, resulting in improved generalization and long-term recommendation performance.

\section{Agentic Replacement: Agent(s) as Recommender}\label{sec:agent_as_recommender}

\subsection{Definition and Scope}\label{subsec:agent_rec_def}

{The paradigm \emph{agent as recommender} refers to systems in which an intelligent agent directly assumes the role of the recommender rather than merely supporting a conventional recommendation pipeline. In such systems, the agent is the end-to-end decision maker that (i) interprets user intent, (ii) maintains user state over time, (iii) selects and executes actions that may include tool invocation and external information gathering, and (iv) produces the final recommendation and natural-language response. The distinguishing feature is \emph{decision authority}: classical recommendation models, if present, are invoked as tools rather than acting as the primary controller \cite{huang2024interecagent,wang2024recmind}.}

{
To systematize the diverse methodologies in agent as recommender, we categorize existing works into two primary architectural paradigms: \textbf{Single-Agent} and \textbf{Multi-Agent}. While both paradigms share the fundamental components of agents (such as Profile, Memory, Tool-Using, Workflow, and Optimization), they diverge fundamentally in how these components are orchestrated and how decision-making authority is distributed.
}

\textbf{Single-Agent based Recommender} functions as a centralized, monolithic decision-maker. In this paradigm, a single agent acts as the unified ``brain,'' responsible for the entire lifecycle of the recommendation process—from interpreting user intent and managing internal memory to invoking tools and generating the final response. The core challenge here lies in enhancing the individual agent's capabilities to handle complex reasoning chains without losing context or coherence. As detailed in \S~\ref{subsec:agent_rec_components}, our discussion focuses on how individual modules (e.g., Profile, Memory) are internalized within a single entity to achieve proactivity and adaptivity.

\textbf{Multi-Agent based Recommender}, in contrast, operates as a decentralized collaborative ecosystem. Here, the recommendation task is decomposed into sub-problems assigned to specialized agents (e.g., a Planner, a Critic, or a Domain Expert). The system's intelligence emerges not from a single model's depth but from the interaction between agents. This paradigm addresses the limitations of single-agent contexts by distributing cognitive load and introducing mechanisms like debate and role-playing. Consequently, in \S~\ref{subsec:agent_rec_multi}, our focus shifts from internal modules to inter-agent dynamics, specifically emphasizing \textit{Communication} protocols and collective \textit{Optimization}.

This taxonomy allows us to dissect how the five core components evolve: roughly speaking, what manifests as an internal \textit{Workflow} in a single agent often scales into a \textit{Communication} protocol in a multi-agent system; similarly, internal \textit{Self-Reflection} evolves into peer-to-peer \textit{Feedback}. 
{
In the following, we first elaborate on \textit{single-agent based recommender}, discussing how the five core components are internalized within a single agent to achieve proactivity and adaptivity for recommendation (\S \ref{subsec:agent_rec_components}). 
Then, we shift from internal modules to inter-agent dynamics and focus on \textit{Multi-Agent based Recommender} (\S \ref{subsec:agent_rec_multi}), specifically emphasizing \textit{Communication} protocols and collective \textit{Optimization}. 
}

\subsection{Single-Agent Recommenders (LoA L4)}\label{subsec:agent_rec_components}

\textbf{Modeling recommender as a single agent substantially upgrades their high-level capabilities beyond what traditional model-centric pipelines can offer.} 
1) As a unified decision-making entity, a single agent enables \textit{proactivity}, transitioning the system from purely reactive responses to actively inferring needs, initiating clarifications, and planning ahead within a session. 2) It also enhances \textit{context awareness} by maintaining a coherent internal state: while the underlying model is constrained by a finite context window, the agent can leverage explicit memory or state tracking to preserve user preferences and interaction history across turns. 
3) Moreover, the single-agent paradigm improves \textit{interaction flexibility}, supporting multi-turn, natural-language preference elicitation and iterative refinement rather than single-shot prediction. 
4) Finally, a single agent introduces \textit{stronger adaptivity}: it can continuously update its internal beliefs, adjust its recommendation strategy based on ongoing feedback, and self-correct within a session, instead of acting as a static mapping from input to output. Together, these improvements make the single-agent recommender more conversational, contextually grounded, and behaviorally adaptive than traditional recommender systems.

{To systematize the literature, we follow a reusable decomposition into five components: profile (\S \ref{subsubsec:agent_rec_profile}), memory (\S \ref{subsubsec:agent_rec_memory}), tool-using (\S \ref{subsubsec:agent_rec_tools}), workflow (\S \ref{subsubsec:agent_rec_workflow}), and optimization (\S \ref{subsubsec:agent_rec_optimization}). This decomposition is practical because it enables comparison between systems that differ in task setting (\eg conversational versus sequential recommendation) but share common control-loop structures.}

\subsubsection{\textbf{Profile: Persistent Preference and Constraint Modeling}}\label{subsubsec:agent_rec_profile}

{The \emph{profile} is a persistent representation capturing long-term preferences, constraints, and relatively stable aspects of user intent. Unlike classical recommenders where profiles are often implicit (encoded in latent embeddings), agentic recommenders frequently maintain explicit textual or structured profiles to support interpretability and controllability \cite{huang2024interecagent,tang2025interactive}. Profiles can be constructed by summarizing interaction histories into preference statements, maintaining structured constraint lists (budget, time, category exclusions), or extracting higher-level values that explain the user’s choices.}

{
In agentic recommender systems, the profile serves as the foundation for modeling both users as autonomous entities with distinct goals, preferences, and behavioral patterns.
For instance, in InteRecAgent~\cite{huang2024interecagent}, the profile module maintains a structured user representation comprising three facets (i.e., like, dislike, and expect) which are dynamically synthesized by the LLM from conversational history to capture both long-term preferences and short-term intentions for tool invocation.
Instruct$^2$Agent~\cite{xu2025iagent}  maintains an instruction-aware profile updated with round-wise user feedback and a dynamic extractor that derives task-specific preferences under current instructions, enabling per-user optimization decoupled from other users.
}

\subsubsection{\textbf{Memory: Working, Episodic, and Semantic State}}\label{subsubsec:agent_rec_memory}
{Memory is a core component that supports contextual reasoning and provides grounding evidence that constrains the agent’s outputs. A useful abstraction distinguishes: (i) \textbf{\textit{working memory}} (short-term conversational context), (ii) \emph{\textbf{episodic memory}} (retrievable records of prior interactions and feedback), and (iii) \emph{\textbf{semantic memory}} (abstracted preference facts and world knowledge) \cite{wang2024recmind,liu2025agentcf++}.} 
Memory can also take different forms, including \textbf{textual} (explicit) representations, \textbf{parametric} (implicit) representations, or none.

\begin{table}[!t]
\caption{Taxonomy of memory mechanisms and representative works in agentic recommender systems.}
\label{tab:memory}
\small
\setlength{\tabcolsep}{4pt}
\begin{tabularx}{\columnwidth}{p{0.18\columnwidth} X p{0.28\columnwidth}}
\toprule
\textbf{Memory Type} & \textbf{Representative Work} & \textbf{High-level Capabilities} \\
\midrule
\textbf{Working Memory} 
& InteRecAgent~\cite{huang2024interecagent}, ToolRec~\cite{zhao2024toolrec}, AgentRecBench~\cite{shang2025agentrecbench}, InstructAgent~\cite{xu2025iagent}, AgentDR~\cite{yang2025agentdr}, RuleAgent~\cite{wang2025ruleagent}, PUMA~\cite{cai2025large}, VibeMus~\cite{guo2025vibemus}, Sunnie~\cite{wu2024sunnie}
& \multirow{3}{=}{\newline\textbf{Context Awareness:} Memory\newline
\textbf{Interaction Flexibility:} Strengthen multi-turn interaction\newline
\textbf{Adaptivity:} Continuous evolving} \\
\cmidrule{1-2}
\textbf{Episodic Memory} 
& PUMA~\cite{cai2025large}
& \\
\cmidrule{1-2}
\textbf{Semantic Memory} 
& RecMind~\cite{wang2024recmind}, AFL~\cite{cai2025agentic}, RecAI~\cite{lian2024recai}, InstructAgent~\cite{xu2025iagent}, AgentDR~\cite{yang2025agentdr}, MemRec~\cite{chen2026memrec}
& \\
\bottomrule
\end{tabularx}
\end{table}

{
We summarize the three types of memory and how it achieves different agentic capabilities in Table~\ref{tab:memory}, while we give some concrete examples as follows. 
In InteRecAgent~\cite{huang2024interecagent}, the memory module is realized as a Candidate Bus, which maintains current item candidates separately from the prompt by combining a data bus—initialized at each conversation turn with all items or user-specified candidates and updated after each tool execution—and a tracker that logs each tool’s input, output, and execution status, enabling sequential streaming of candidates through multiple tools and supporting the reflection mechanism for judgment.
RecMind~\cite{wang2024recmind} defines a Memory component split into: Personalized Memory, which stores individual user data (e.g., their ratings or reviews), and World Knowledge, which stores item metadata (domain‑specific) and real‑time information via web search.  
BiLLP~\cite{shi2024planner} designed dedicated working memory modules for each of its key components (e.g., actor, critic, and planner) to store past experiences, with memory updated after each step.
The memory of \cite{yu2025intelligent_agent} selectively stores relevant item features, user reviews, and interaction history to support accurate and personalized recommendations continuously.
}

{
\subsubsection{\textbf{Tool-Using: Expanding the Action Space.}}\label{subsubsec:agent_rec_tools}
A key characteristic of agentic systems is their ability to augment reasoning through external tools, enabling agents to dynamically interact with APIs, databases, search engines, or specialized modules to enhance their recommendation capabilities. We categorize the tools employed by existing agents into four main types:
(1) \textbf{RecTools}: Tools for retrieval, ranking, or classical recommender system engines, which form the core mechanism for filtering relevant candidates and predicting user preferences;
(2) \textbf{External knowledge}: Tools that access knowledge bases or search engines beyond the recommendation system, supporting responses to user queries that require real-time or external information~\cite{guo2024knowledge};
(3)
\textbf{Attribute tools}: Schema-aware or faceted filters for attribute-oriented retrieval, allowing fine-grained control over item selection;
(4)
\textbf{Multimodal tools}: Tools designed to handle additional modalities such as images, audio, or code, extending the agent’s capability beyond textual information.}

{
For instance, in InteRecAgent~\cite{huang2024interecagent}, retrieval and ranking tools serve as the core recommendation engines by identifying relevant candidates and estimating user preferences, while information query tools allow the agent to access external knowledge bases or databases via SQL and search engines to answer user inquiries beyond standard recommendation tasks.
RecMind~\cite{wang2024recmind} mainly integrates three types of tools to access and process external knowledge: a Database Tool that converts natural language queries into SQL to retrieve in-domain knowledge such as user reviews or item metadata; a Web Search Tool for fetching real-time information from the internet as part of world knowledge; and a Text Summarization Tool to process and condense the retrieved information.
In the multi-agent collaboration framework proposed by MACRec~\cite{macrec}, the Analyst is granted access to two tools to assist in analysis: an information database to get the user profile and item attributes and an interaction retriever to get the user/item interaction history.
ToolRec~\cite{zhao2024toolrec} introduces attribute-oriented retrieval and ranking tools: retrieval tools first return candidate items based on specified attribute patterns and size constraints, while ranking tools use LLMs with instruction templates to order these candidates according to user history and attribute relevance, effectively capturing users’ latent intent without training separate models for each attribute.
\cite{shi2024planner} define a Category Analysis Tool which can identify a list of categories associated with each legal action and conduct a statistical analysis on the user’s
viewing history.
AgentDR~\cite{yang2025agentdr} bridges LLM reasoning with scalable rec tools by delegating full-catalog ranking to traditional recommenders while using the agent to integrate multiple model outputs (personalized tool suitability) and inject commonsense relational reasoning over substitutes/complements grounded in the user’s history—mitigating hallucination and keeping the system scalable.
}

\begin{table}[!t]
\caption{Taxonomy of workflow controllers in agentic recommendation systems.}
\label{tab:workflow_controllers}
\vspace{-8pt}
\small
\setlength{\tabcolsep}{4pt}
\begin{tabularx}{\columnwidth}{p{0.18\columnwidth} X p{0.28\columnwidth}}
\toprule
\textbf{Workflow Type} & \textbf{Representative Work} & \textbf{High-level capabilities} \\
\midrule
\textbf{ReAct} 
& RecMind~\cite{wang2024recmind}, ToolRec~\cite{zhao2024toolrec}, AgentRecBench~\cite{shang2025agentrecbench}, PUMA~\cite{cai2025large}, WeMusic-Agent\cite{bi2025wemusic}
& \multirow{3}{=}{\newline \textbf{Proactivity}: proactive \& reactive,\newline \newline
\textbf{Interaction Flexibility}: Multi-turn,\newline \newline
\textbf{Adaptivity}: continuous evolving} \\
\cmidrule{1-2}
\textbf{Plan-then-Execute} 
& InteRecAgent~\cite{huang2024interecagent}, AFL~\cite{cai2025agentic}, RecAI~\cite{lian2024recai}, AgentRecBench~\cite{shang2025agentrecbench}, InstructAgent~\cite{xu2025iagent}, CogRec~\cite{hu2025cogrec}, ScienceDB AI~\cite{long2026sciencedb}, AMEM4Rec~\cite{nguyen2026amem4rec}
& \\
\cmidrule{1-2}
\textbf{Reflex}
& AgentRecBench~\cite{shang2025agentrecbench}, AgentDR~\cite{yang2025agentdr}, RuleAgent~\cite{wang2025ruleagent}
& \\
\bottomrule
\end{tabularx}
\end{table} 

\subsubsection{\textbf{Workflow Controllers: From Reasoning--Acting to Reflection}}\label{subsubsec:agent_rec_workflow}

\begin{figure}[t]
\setlength{\abovecaptionskip}{-0.0cm}  
\setlength{\belowcaptionskip}{-0.0cm} 
    \centering
    \includegraphics[width=1\linewidth]{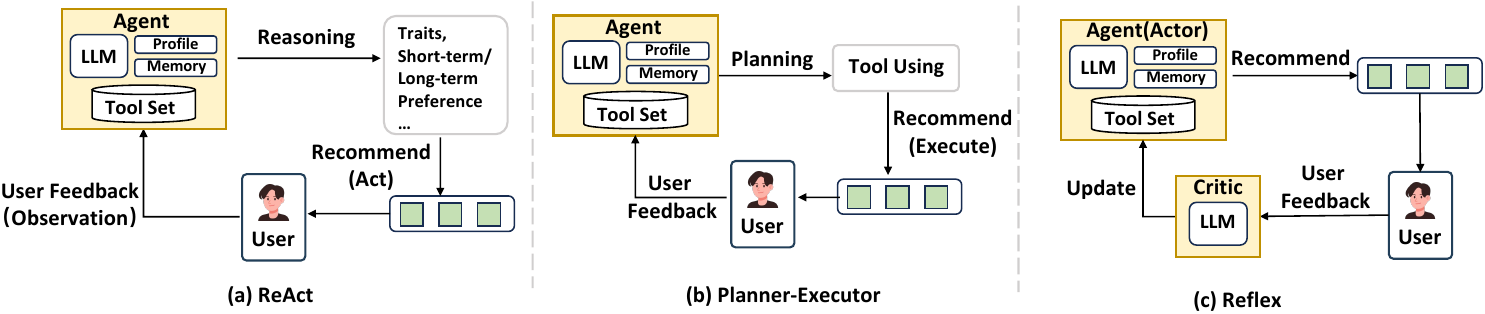}
    \caption{Three paradigms of workflows of single-agent recommenders.}
    \label{fig:workflow}
\end{figure}

{The workflow controller specifies how the agent sequences reasoning steps and actions. As illustrated in Figure~\ref{fig:workflow}, three recurring patterns are prominent.}
Table~\ref{tab:workflow_controllers} summarizes these workflow paradigms, along with representative methods and their associated high-level capabilities. 
Overall, ReAct emphasizes flexible interaction and incremental reasoning, Plan-then-Execute improves controllability and structured decision-making, while Reflex enables continuous policy refinement through self-evaluation. 
These complementary designs collectively support key agentic capabilities, including proactivity, multi-turn interaction flexibility, and adaptive evolution.

\noindent$\bullet\quad$\textbf{ReAct (interleaved reasoning and acting).}
{In this pattern, the agent alternates between intermediate reasoning and tool invocation, supporting incremental refinement of candidates and constraints \cite{yao2022react}. This is effective for open-ended requests but can compound errors if early tool calls are misguided.} 
{
A large portion of existing work lies in this workflow. MoRE~\cite{more} introduces a reflection-based framework for sequential recommendation, where an LLM dynamically selects and combines multiple reflectors to iteratively reconsider and correct its recommendation decisions, mitigating biases and instability caused by a single reasoning path. DRDT~\cite{drdt} uses dynamic reflection with divergent thinking within a retriever–reranker framework to iteratively probe, critique, and tailor LLM reasoning to better model sequential user preferences over time.
R4ec~\cite{gu2025r} establishes a reasoning–reflection–refinement loop where an actor model proposes recommendations and a reflection model evaluates and feeds back corrections, enabling deliberate System-2-like thinking to improve recommendation accuracy. Re2LLM~\cite{re2llm} guides LLMs to self-reflect on recommendation errors to build a hint knowledge base and trains a lightweight agent to select useful hints that correct reasoning, resulting in more accurate session-based recommendations without fine-tuning the LLM. SRLF~\cite{srlf} introduces a set-wise reflective learning framework that uses an LLM agent to assess entire candidate item sets, detect mismatches with user feedback, and iteratively refine both user preferences and item semantics through a closed-loop reflection process, capturing richer inter-item relationships for better sequential recommendation. 
}

\noindent$\bullet\quad$\textbf{Plan-then-Execute.}
{Here, a planner decomposes the user request into subgoals (\eg constraint inference, candidate retrieval, verification) and an executor carries out tool calls under this plan. This improves controllability because plans can be inspected and constrained \cite{huang2024interecagent,yu2025thought}.} 
For example, InteRecAgent~\cite{huang2024interecagent} adopts a plan-first execution workflow consisting of two stages: in the first stage, the LLM generates a complete tool-utilizing plan based on the user query; in the second stage, the system executes each operation sequentially according to the plan, without invoking the LLM at every step. 
RecMind~\cite{wang2024recmind} introduces a Self-Inspiring (SI) planning method that leverages all previously explored reasoning branches to generate each step, enabling more comprehensive and multi-perspective reasoning for recommendation tasks than traditional single-path approaches like CoT~\cite{wei2022cot} and ToT~\cite{yao2023tot}. 
Instruct$^2$Agent~\cite{xu2025iagent} follows plan-then-execute (parser $\rightarrow$ knowledge/tool use $\rightarrow$ reranker) with a reflex-style self-reflection that regenerates constrained lists when inconsistencies are detected.

\noindent$\bullet\quad${
\textbf{Reflex}: Self‑critique loop or judge–checker pattern~\cite{wu2025starec,wu2025personalized}.
BiLLP~\citet{shi2024planner}  
follow the actor-critic framework and initialize the actor and critic as using LLM, 
to provide personalized recommendations, where Critic assesses the user’s
current satisfaction level (action advantage value) and updates
the policy of Actor to enhance personalized recommendations.
T-PRA~\cite{wang2025t-pra} adopts an actor–critic framework to balance the trade-off between user satisfaction and user
interest exploration in proactive conversational recommendation.
AgentDR~\cite{yang2025agentdr}follows plan-then-execute with a reflex-style check: the agent plans which base recommenders to consult, aggregates their candidates, reasons about item-item substitutes/complements, and re-scores candidates before output—acting as a controller atop conventional rankers rather than a monolithic generator.
}

\subsubsection{\textbf{Optimization and Adaptation}}\label{subsubsec:agent_rec_optimization}

{
To continuously enhance the decision-making capabilities, recommendation agents must perform optimization over time—refining their internal policies and strategies through iterative learning.
Such optimization processes can be broadly categorized into two complementary paradigms:
(1) \textbf{Self reflection}, where the agent evaluates and improves upon its own reasoning and recommendations; and
(2) \textbf{Feedback reflection}, where the agent adapts based on external signals from the environment or user feedback.
Together, these mechanisms enable agents to move beyond static policy learning toward agentic optimization, characterized by proactive self-improvement and adaptive alignment with long-term user goals.
}

\begin{figure}[t]
\setlength{\abovecaptionskip}{-0.0cm}  
\setlength{\belowcaptionskip}{-0.0cm} 
    \centering
    \includegraphics[width=0.8\linewidth]{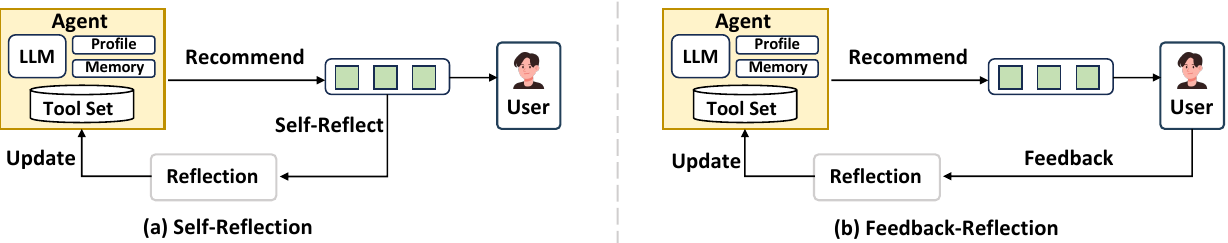}
    \caption{Two paradigms of optimization in single-agent recommenders.}
    \label{fig:optimization}
\end{figure}

{
\textbf{Self-Reflection:}
Self-reflection focuses on internal critique and self-improvement loops, where the agent evaluates the quality of its own decisions and refines its policy accordingly.
For instance, T-PRA~\cite{wang2025t-pra} introduces a critic module that generates structured feedback on recommendation outcomes. This feedback is then leveraged via Direct Preference Optimization (DPO) to jointly refine both the actor and advisor components, improving alignment with long-term proactive recommendation objectives.

\textbf{Feedback-Reflection:} In contrast, feedback-reflection leverages external signals from user interaction or environment feedback to guide optimization.
ECPO~\cite{feng2025expectation} exemplifies this paradigm by modeling user expectations and confirmations at each conversational turn. It identifies sources of user dissatisfaction and performs fine-grained, turn-level preference updates. This design improves multi-turn interaction quality while avoiding the high sampling cost associated with prior methods such as MTPO.
}

\subsection{Multi-Agent Recommenders (LoA L5)}\label{subsec:agent_rec_multi}

\textbf{Beyond the capabilities enabled by a single agent, multi-agent recommenders introduce an additional layer of enhancement by distributing decision-making across multiple specialized agents.} 
1) In terms of proactivity, multi-agent systems move from individual proactive reasoning to \textit{collective proactivity}, where different agents—such as planners, critics, retrievers, and preference elicitors—jointly anticipate user needs and cross-check each other’s actions, leading to more reliable proactive behaviors than a single agent. 
2) For context awareness, multi-agent architectures alleviate the limitations of a single agent’s context window by allowing each agent to maintain its own perspective or memory, resulting in distributed memory that captures user intent, item evidence, and environmental constraints more completely. 
3) Regarding interaction flexibility, multi-agent setups transform single-agent multi-turn interaction into multi-party collaborative interaction: agents can debate, critique, refine, or negotiate with one another before presenting results to the user, enabling richer reasoning patterns than any single agent can achieve. 
4) Finally, in terms of adaptivity, multi-agent systems support continuous evolving behavior not only through user feedback but also through internal self-correction loops—where agents iteratively refine each other’s outputs—making the system more adaptive and stable compared to a single adaptive policy. 
Overall, while single-agent systems enhance recommendation along the four dimensions within one coherent policy, multi-agent systems amplify these capabilities through structured cooperation, role specialization, and self-consistency mechanisms.

\subsubsection{\textbf{Roles and Coordination Protocols}}\label{subsubsec:agent_rec_coordination}
Common role sets include manager/planner agents, retrievers, rankers, analyzers, critics, verifiers, and safety monitors. Coordination protocols vary: manager--worker decomposition \cite{wang2024macrec, xia2026multi, zhang2026llms,nie2024hybrid,portugal2024agentic,agarwal2024multistage}, debate--judge selection \cite{fang2024multi,ma2025agentrec}, role-based trustworthy conversational protocols \cite{hui2025matcha}, and negotiation-based approaches for multi-stakeholder recommendation \cite{banerjee2025collab, dixit2026pcn}. Hierarchical agent structures are also common in route and itinerary recommendation, where feasibility constraints naturally require staged reasoning and verification.

\subsubsection{\textbf{Communication}} Communication is a key mechanism enabling collaboration and coordination in multi-agent recommendation systems.
By exchanging information about user context, preferences, and reasoning outcomes, agents can jointly construct more accurate and consistent recommendations.
Such inter-agent communication fosters cooperative behaviors, mitigates information asymmetry, and supports the emergence of specialized agent roles for complex recommendation tasks.
MacRec~\cite{macrec} introduces a multi-agent collaboration framework, including the Manager, User/Item Analyst, Reflector, Searcher, and Task Interpreter, to enhance recommendation tasks.
MAS4POI~\cite{wu2025mas4poi} is a multi-agent framework (comprising agents such as DataAgent, Manager, Analyst, and Navigator) that collaboratively produces next point-of-interest (POI) recommendations. In each task, one or more agents are selected to perform the corresponding operation, and all agents are coordinated and their information aggregated through the manager agent.
MACRS~\cite{fang2024multi} decomposes the Conversational Recommender System (CRS) workflow into four LLM-based agents that plan dialogue acts, independently generate candidate responses, and then select a final response via a judge module. Beyond the policy side, MACRS also provides a more realistic user simulator by masking the target item with a keyword-level profile to reduce direct leakage and overly explicit hints.
In ARAG~\cite{maragheh2025arag}, agents communicate via staged hand-offs rather than free-form chat: understanding $\rightarrow$ alignment-scoring $\rightarrow$ evidence compression $\rightarrow$ ranking. This act-structured passing stabilizes dialogue flow and makes decisions diagnosable.
In RecBot~\cite{tang2025interactive}, communication between the Parser and Planner agents follows a structured hand-off protocol, where the Parser conveys normalized user intents to the Planner through explicit JSON-style messages, ensuring transparent coordination and preventing semantic drift during multi-turn command execution.
In Collab-REC~\cite{banerjee2025collab} , communication proceeds via moderated rounds: agents submit act-conditioned proposals; the moderator feeds back rejections and penalties (e.g., repeated or spurious cities) and requests revisions, yielding negotiated consensus rather than free-form chat.
TAIRA~\cite{yu2025thought} coordinates a Manager–Executors pipeline via structured hand-offs: the Manager selects a thought pattern, decomposes the user request into sub-tasks, and dispatches them to specialized Executors, who return evidence and partial results for aggregation—reducing semantic drift compared with free-form multi-agent chat.
In MATCHA~\cite{hui2025matcha}, inter-agent communication follows a structured role-based protocol, where the intent agent passes normalized user goals to the generator and ranker, and downstream explanation and safeguard agents exchange rationale and risk annotations to ensure coherent, transparent, and trustworthy collaboration.

\subsubsection{\textbf{Optimization}}\label{subsubsec:multi-agent-optimization}
Optimization in multi-agent recommendation systems focuses on improving collective decision quality without relying on gradient-based parameter updates.
Through mechanisms such as structured coordination, negotiation, and self-distillation, agents can iteratively refine their reasoning and cooperation strategies~\cite{acharya2026group, li2026recnet, wu2026internalizing, wang2026self, wang2026momorec, zhu2025llm_based_conv}.
Rather than relying on gradient-based updates, MACRS~\cite{fang2024multi} mainly optimizes the decision quality through structured coordination (parallel candidate generation followed by selection) and by adopting a more realistic user-simulation protocol.
ARAG~\cite{maragheh2025arag} optimizes recommendation quality by coordinating agents for consistency-aware evidence refinement, enabling the system to iteratively filter irrelevant retrievals and enhance ranking accuracy without retraining the backbone LLM.
RecBot~\cite{tang2025interactive} optimizes its dual-agent policy through simulation-driven distillation, where high-fidelity trajectories from a teacher model (e.g., GPT-4.1) are distilled into a lightweight student agent, enabling efficient online deployment while preserving reasoning and planning competence.
Collab-REC~\cite{banerjee2025collab} optimizes decision quality without fine-tuning LLMs by combining multi-round negotiation with a penalty-aware scoring function that rewards agent success while penalizing repetition and hallucinated entries—explicitly steering the system toward balanced, diverse lists across stakeholders.
TAIRA~\cite{yu2025thought} optimizes agent behavior through thought-pattern distillation, a self-improving process that abstracts successful reasoning traces into reusable planning templates, enabling more consistent task decomposition and tool use without retraining the underlying LLM.

\subsection{Summary}\label{subsec:agent_rec_summary}

Agent-as-recommender systems represent the most direct realization of autonomy in recommendation: the agent becomes the controller and the decision maker. The literature indicates a progression from single-agent interactive controllers, to tool-learning and reflective workflows, and finally to multi-agent orchestration for robustness, negotiation, and complex domains. 
From agentic augmentation (\ie agent-assisted recommender) to agentic replacement (\ie agent as recommender), recommender systems are more proactive, have stronger context awareness, are more flexible in interactions, and more adaptive (\cf \S \ref{subsec:agent_rec_components} and \S \ref{subsec:agent_rec_multi}).

\section{Agentic Simulation: Data Synthesis, User Simulation, and Environment Simulation}
\label{sec:simulation}

In this survey, \emph{simulation} refers to using one or more agents to approximate parts of the recommendation ecosystem that are expensive, unsafe, or impossible to observe directly.
This includes generating synthetic training data, simulating user feedback to train or test recommenders, and simulating closed-loop environments where recommender policies interact with populations over time.

\subsection{Why do we need simulation?}\label{subsuc:simulation_motivation}
Across the literature, simulation is used for four recurring motivations: 
1) \textbf{Training without expensive interaction.} Simulators provide behavioral signals when online exploration is infeasible (cold-start, privacy constraints, or limited traffic). 
2) \textbf{Counterfactual and stress testing.} Simulators enable controlled tests of robustness, fairness, and safety properties that are difficult to measure from logs. 
3) \textbf{Data synthesis and augmentation.} Synthetic dialogues, profiles, and feedback can reduce sparsity and broaden coverage. 
4) \textbf{Ecosystem and feedback-loop analysis.} Multi-agent simulations can model long-term dynamics such as filter bubbles and echo chambers.

\subsection{A taxonomy of simulation types}
We propose a simple taxonomy that aligns with autonomy and the “what is being simulated” question.

\begin{figure}[t]
\setlength{\abovecaptionskip}{-0.0cm}  
\setlength{\belowcaptionskip}{-0.0cm} 
    \centering
    \includegraphics[width=1\linewidth]{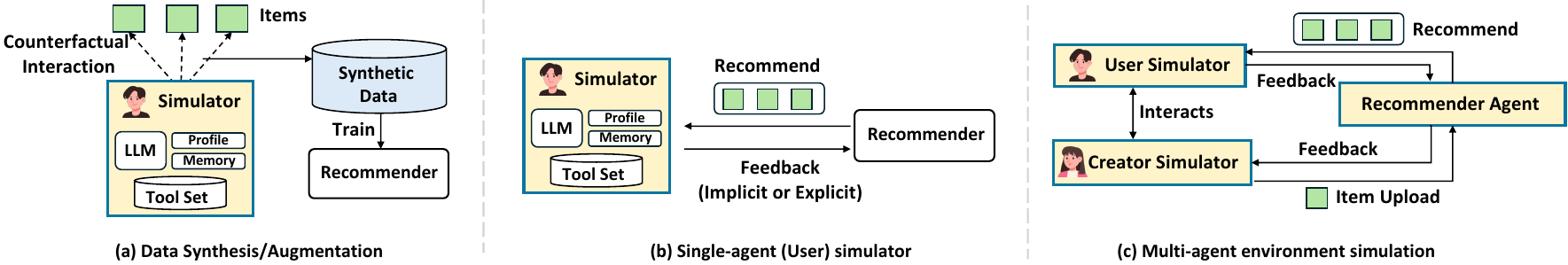}
    \caption{Three taxonomies of agentic simulation: data synthesis, user (single-agent) simulation, and environment (multi-agent) simulation.}
    \label{fig:simulation}
\end{figure}

\subsubsection{\textbf{(1) Data synthesis / data augmentation}}\label{subsubsec:simulation-data-synthesis}
Data synthesis uses agents to generate artifacts that support training or evaluation: synthetic conversations, item descriptions, preference rationales, or structured user profiles.
TalkPlayData~2 exemplifies an agentic synthetic pipeline for multi-modal conversational music recommendation, demonstrating how agentic generation can create richer training signals than template-based augmentation \cite{choi2025talkplaydata}. 
For cold-start settings, simulator-style generation can replace missing interaction histories with plausible synthetic signals: a large language model simulator for cold-start recommendation explicitly targets the absence of real user interaction data \cite{huang2025large}.  
LAUS~\cite{park2025llm} leverages large language models to simulate user interactions, enabling the training of news recommenders without relying on large-scale real user data.
For reinforcement learning-based recommender systems, \citet{zhang2025llm} propose an interpretable user simulator that combines LLM-based preference reasoning with statistical behavior modeling to generate high-fidelity training data, thereby improving recommendation training.

\subsubsection{\textbf{(2) User simulation (single-agent simulators)}}\label{subsubsec:simulation-single-agent-simulator}
User simulation models user feedback (clicks, ratings, critiques, dialogue acts) conditioned on context.
This can appear as a single LLM agent that plays the user role, or as structured simulators augmented by LLM reasoning. 
A key example is RecAgent, which frames simulation as a paradigm for recommender systems and highlights how LLM agents can generate interactive trajectories rather than static labels \cite{wang2023recagent}.
User Behavior Simulation with LLM-based Agents provides a systematic formulation and is especially relevant because it positions LLM-agent simulators as a way to reproduce behavioral dynamics at scale \cite{wang2025user}, requiring alignment with real-world data distributions.

\paragraph{\textbf{Design axes for user simulators.}}
User simulators can be compared along:
(i) \textbf{fidelity} (does the simulator match real distributions and causal responses?),
(ii) \textbf{controllability} (can we steer demographics, noise, and drift?),
(iii) \textbf{observability} (does the simulator output only implicit feedback or also explicit rationales?),
and (iv) \textbf{calibration} (does the simulator’s actions reflect real behaviors?).
These axes should be evaluated explicitly rather than assumed, because simulator errors can lead to “training on simulator artifacts.” 
In the following, we discuss the existing literature from these design axes, respectively.

(i) \emph{Simulation fidelity}. To ensure high simulation \emph{fidelity} aligned with real-world data distributions, user simulators primarily focus on two aspects: {\textit{profile construction}} and {\textit{memory design}}.
\begin{itemize}
    \item \textbf{\textit{Profile construction}}. Most existing user simulators construct personalized user profiles from real-world interaction data~\cite{wanyan2025temporal,liu2025diagnostic}. For example, RecAgent~\cite{recagent} and Lusifer~\cite{ebrat2025lusifer} build basic profiles using demographic and attribute information such as gender, age, traits, and interests, while Agent4Rec~\cite{zhang2024agent4rec} derives social traits (e.g., activity, conformity, and diversity) from datasets like MovieLens-1M. In addition, several methods, including RecAgent~\cite{wang2023recagent} and Agent4Rec~\cite{zhang2024agent4rec}, leverage LLMs for generative profiling, such as summarizing representative behavioral roles, personalized preferences, and rating patterns.
    \item \textbf{\textit{Memory design}}. Another key component to ensure fidelity is memory. It allows the agent to recall past actions, feedback, and outcomes, thereby supporting sequential decision-making and adaptive behavior in long-term recommendation simulations. 
    Beyond saving long- and short-term interests, memory mechanisms in user simulators are designed to capture richer aspects of user behavior. For instance, RecAgent~\cite{recagent} introduces sensory memory, which directly interacts with the environment and transforms raw observations into concise representations. Agent4Rec~\cite{zhang2024agent4rec} proposes factual memory, which encodes user interaction behaviors, and emotional memory, which captures the psychological states induced by these interactions.
\end{itemize}

(ii) \emph{Simulator controllability}. A controllable user simulator adds an important dimension: \emph{controllability} over behavior and dynamics, which is critical for evaluation and for testing robustness to demographic or preference shifts \cite{zhu2025llm,wei2025mirroring,zhu2024reliable}. 
\citet{zhu2025llm} proposes a plugin-based, controllable and human-in-the-loop LLM user simulator (CSHI), where controllability is enabled by a modular plugin manager for stage-wise behavior control, explicit user profile manipulation, and human intervention.
Additionally, controllability can be achieved by adjusting the simulator’s memory decay rate to model interest drift~\cite{wang2023recagent, ye2025creagent}, and by tuning the sampling distribution of user profiles to simulate different demographic compositions~\cite{ye2025creagent, zhang2024agent4rec}.

(iii) \emph{Simulation observability}. Beyond outputting implicit feedback (e.g., like), improving \emph{observability} of the recommendation process is another key ability of user simulator. \citet{zhang2025llm} propose an interpretable user simulator that integrates LLM-based preference reasoning with statistical behavior modeling to generate high-fidelity training data, thereby improving reinforcement learning–based recommender systems.
For domain-specific simulators, LLM-as-user simulation for news recommendation targets training without real interactions \cite{park2025llm}, while Lusifer provides an LLM-based simulated feedback environment for online recommender systems \cite{ebrat2025lusifer}.

(iv) \textit{Behavior calibration}. 
Whether the simulator’s behavior reflects real user behavior is a critical aspect of simulator design. Existing research primarily models two types of actions: (1) user-system actions and (2) user-user actions.

\begin{itemize}
    \item 
\textbf{User-system action.} In non-conversational user simulation, implicit signals (e.g., like, click)~\cite{zhang2024agent4rec, park2025llm, zhang2025llm, huang2025large, bougie2026alignuser} and explicit signals (e.g., comments, reviews)~\cite{recagent} constitute the two primary types of user-to-item actions in user simulation. To implicitly model,
Agent4Rec~\cite{zhang2024agent4rec} divides user page-by-page browsing actions into taste-driven and emotion-driven types: (1) taste-driven actions include viewing, rating, and expressing post-viewing feelings towards items, capturing users’ immediate preferences during page-by-page recommendations; (2) emotion-driven actions (i.e. existing the session or rating the recommender system) capture user affective states and reflect how satisfaction and fatigue shape their decision to continue or quit. Beyond user feedback simulation, some works~\cite{ye2025creagent, jin2025recinter} model item providers (e.g., creators or merchants), enabling simulators to update item attributes or generate new items to capture supply-side dynamics.
In conversational recommender systems, users typically interact with the system through the generative capabilities of LLMs~\cite{chen2025recusersim, feng2025expectation}, generating fluent and contextually relevant responses to express preference.

\item \textbf{User-user action.} 
To capture this broader social dimension, recent works~\cite{recagent} introduce conversational user simulators that equip LLM-based agents with the ability to interact with other user simulators, enabling more realistic and socially grounded simulations.
RecAgent~\cite{recagent} not only enables users to perform conventional recommender system actions (e.g., searching, browsing), but also allows agents to engage in one-to-one chatting and one-to-many broadcasting.

\end{itemize}

\subsubsection{\textbf{(3) Closed-loop multi-agent environment simulation}}\label{subsubsec:simulation-closed-loop-multi-agent-environment}
The most ambitious form of simulation uses multiple agents to model populations and platform mechanisms, enabling closed-loop experiments over long horizons.
RecoWorld provides simulated environments for agentic recommender systems, pushing toward standardized testbeds where policies can be compared under controlled dynamics \cite{liu2025recoworld}.
Filter bubble simulation with LLM agents demonstrates that such environments can capture emergent societal effects (e.g., self-reinforcing exposure patterns) that are central to multi-objective evaluation \cite{sukiennik2025simulating}. 
GGBond extends this direction with a graph-based AI-agent society, highlighting the need to represent social structure and influence in ecosystem simulations \cite{zhong2025ggbond}. 
\citet{cai2025agentic} proposes an agentic feedback loop framework that explicitly models the iterative interaction between a recommendation agent and a user agent, allowing both to co-evolve and improve through mutual feedback.
Beyond user simulation, recent works extend the paradigm to multi-agent environments by modeling item providers  (e.g., merchants, content creators). In this setting, simulators can act not only as a user model that mimics user behavior passively responding to recommendations, but also as an active recommendation agent that proactively generates and recommends items to other users. For example, CreAgent~\cite{ye2025creagent} leverages LLMs to simulate a realistic content recommendation platform (i.e., YouTube) involving both users and content creators, enabling long-term offline evaluation of recommender systems with multi-stakeholders. RecInter~\cite{jin2025recinter} introduces merchant agents that can respond to user feedback and dynamically modify item attributes, enabling a more realistic co-evolution of users, items, and providers.

\subsection{Simulation shortcoming and open challenges}
Despite rapid progress, simulation brings substantial risks: 
1) \textbf{Simulator--policy mismatch.}
Agents trained against a simulator may overfit to simulator quirks, leading to inflated offline performance and poor real-world transfer. This is particularly problematic for agentic systems that exploit loopholes in feedback generation. 
2) \textbf{Compounding error in closed loops.}
In closed-loop simulation, small misspecifications in user response can compound over time, producing misleading conclusions about long-term welfare. 
3) \textbf{Evaluation of simulators themselves.}
Simulators must be evaluated as \emph{models}, not just used as tools. We argue that simulator evaluation should be treated as a first-class part of agentic recommender evaluation (Sec.~\ref{sec:evaluation}), including calibration tests, distributional similarity tests, and targeted behavioral probes. 
4) \textbf{Link to evaluation.}
Simulation is simultaneously an \emph{object} of evaluation and a \emph{method} for evaluation; for agentic recommendation, the two are inseparable. We therefore treat simulation-based evaluation and red-teaming as core methodologies in the next section.

\section{Evaluation and Benchmarking}
\label{sec:evaluation}

Agentic recommender systems (ARS) should be evaluated as interactive systems rather than only as static rankers. In a conventional top-$N$ setting, the main evaluation object is a ranked list compared with held-out interactions. In an ARS, the object is often a trajectory that includes user inputs, internal state, memory reads and writes, tool calls, observations, intermediate rationales, final outputs, and feedback. A system can rank relevant items while using unsupported evidence, exposing sensitive profile attributes, or making repeated tool calls; conversely, a fluent dialogue can still fail to recommend useful items. Evaluation therefore has to separate final recommendation utility from the process that produced it.

For an episode with $T$ turns, we write the trajectory as
\[
\tau=(u_1,s_1,a_1,o_1,y_1,\ldots,u_T,s_T,a_T,o_T,y_T),
\]
where $u_t$ is the user input, $s_t$ is the internal or environmental state, $a_t$ is the agent action, $o_t$ is an observation returned by a tool, memory, simulator, or another agent, and $y_t$ is the user-facing output. Classical metrics mainly evaluate the ranked list contained in $y_T$. Agentic evaluation instead requires a vector of measurements,
\[
m(\tau)=(m_{rec},m_{int},m_{gen},m_{ground},m_{tool},m_{mem},m_{safe},m_{cost}),
\]
where the components summarize recommendation utility, interaction quality, generated-output quality, grounding, tool and planning behavior, memory behavior, safety, and cost. The relevant components depend on both the role of the agent and its Level of Autonomy (LoA). L2 systems require retrieval and grounding checks; L3 systems require tool-use diagnostics; L4 systems require plan, memory, and reflection evaluation; and L5 systems require coordination and failure-attribution metrics. Likewise, agent-assisted recommenders evaluate the marginal value of agents around a classical pipeline, agent-as-recommenders evaluate end-to-end agent behavior, and agent-as-simulators evaluate both the simulator and the recommender policies tested with it.

\begin{table*}[!t]
\centering
\footnotesize
\setlength{\tabcolsep}{3.5pt}
\renewcommand{\arraystretch}{1.08}
\caption{Evaluation matrix for agentic recommender systems.}
\vspace{-8pt}
\label{tab:ars-eval-matrix}
\begin{tabularx}{\textwidth}{@{}p{0.13\textwidth}p{0.20\textwidth}p{0.22\textwidth}X@{}}
\toprule
\textbf{H0 target} & \textbf{H1 protocols} & \textbf{H2 purpose} & \textbf{H3 metrics and required context} \\
\midrule
Recommendation outcome & Offline benchmark; simulation; online or A/B test. & Test whether the final item, list, route, or POI set is useful. & HR/Hit, Recall, Precision, NDCG, MRR, MAP, AUC, CTR/CVR. Report split, candidate set, negative sampling, baselines, LLM/backbone version, and repeated runs. \\
Interaction and generated output & CRS benchmark; user study; dialogue simulation; human or LLM judge; reference comparison. & Test preference elicitation, task completion, response quality, and explanation usefulness. & Success@K, task completion, average turns, satisfaction, BLEU/ROUGE/METEOR/BERTScore, clarity, relevance, helpfulness. Report rubric, judge prompt, evidence visibility, and agreement. \\
Grounding and RAG evidence & Retrieval benchmark; claim-level audit; noisy or adversarial evidence test. & Test whether retrieved user, item, or auxiliary evidence supports the recommendation and explanation. & Retrieval Recall@K/NDCG, context relevance, citation accuracy, faithfulness, hallucination and contradiction rate. Report source, granularity, freshness, retrieval budget, and claim segmentation. \\
Simulator validity & Simulator-vs-log comparison; human validation; downstream utility; simulation-to-A/B alignment. & Test whether simulated users, creators, or environments are plausible and useful. & Distributional similarity, rating/click consistency, diversity, leakage rate, reward, liking ratio, treatment-effect correlation. Report calibration data, persona construction, leakage controls, and uncertainty. \\
Agent trace: tools, memory, planning, coordination & Trace logging; functional validation; module ablation; error taxonomy; case study. & Attribute success and failure to agentic components, not only to the final list. & Tool-selection accuracy, argument validity, execution success, plan coverage, recovery success, memory precision/recall, contradiction rate, agreement, time-to-consensus. Report tool schema, memory store, trace format, and agent-role ablations. \\
Safety, privacy, fairness, and deployment & Red teaming; poisoning/jailbreak tests; privacy audit; fairness analysis; efficiency measurement; production monitoring. & Test robustness, policy consistency, privacy preservation, fairness, and operational feasibility. & Attack success, policy violation, privacy leakage, exposure imbalance, diversity, latency, token/inference cost, throughput, timeout/fallback rate, retention/GMV. Report threat model, attack budget, protected/provider groups, hardware/API setup, traffic split, and confidence intervals. \\
\bottomrule
\end{tabularx}
\end{table*}

\subsection{Targets, Protocols, and Metrics}
\label{subsec:evaluation-taxonomy}

A recurring ambiguity in ARS evaluation is the conflation of targets, protocols, and metrics as summarized in Table~\ref{tab:ars-eval-matrix}. A \emph{target} is the object being evaluated: the ranked list, conversation, explanation, RAG evidence, simulator, tool trace, memory update, multi-agent coordination process, or deployed service. A \emph{protocol} is the procedure used to evaluate the target: offline benchmark, simulation, user study, human annotation, LLM-as-judge, adversarial testing, case study, or online experiment. A \emph{metric} is the observable measurement produced by the protocol: NDCG, Recall, Success@K, average turns, BLEU, faithfulness, attack success rate, latency, token cost, or GMV.

The H0--H3 distinction also helps relate evaluation to LoA. At L2, the main additional target beyond ranking is retrieved evidence: whether user, item, or auxiliary knowledge is relevant and used faithfully. At L3, the target expands to tool traces: tool choice, argument validity, execution success, and recovery. At L4, the target includes plan quality, memory updates, and self-correction. At L5, the target includes communication, agreement, and contribution of each specialist agent. This level-wise view prevents a common mismatch in which a paper claims planning, memory, or multi-agent benefits but evaluates only the final ranked list.

Offline evaluation remains useful because it is cheap, reproducible, and compatible with existing recommender baselines. It is not, however, sufficient evidence of agentic behavior. A high NDCG or HR score does not show that the agent selected the right tool, grounded claims in catalog evidence, maintained a consistent profile, asked useful clarifying questions, or satisfied latency constraints. Simulation, user studies, LLM-as-judge protocols, adversarial tests, and online experiments answer different questions and should be treated as complementary. Simulation tests controlled interaction; user studies test perceived utility; LLM judges scale qualitative assessment but require calibration; adversarial evaluation probes robustness; and online experiments provide causal evidence under deployment constraints.

\subsection{Recommendation, Conversation, and Generated Output}
\label{subsec:outcome-eval}

Classical recommender metrics remain the foundation when the target is a final item choice or ranked list. Most agent-assisted and agent-as-recommender papers therefore report HR, Recall, Precision, NDCG, MRR, MAP, AUC, or CTR-like outcomes. InteRecAgent, ToolRec, AgentDR, MADREC, AgenticRAG, R4EC, and VRAgent-R1 retain top-$K$ or ranking metrics to show that agentic workflows do not reduce final-list relevance \cite{huang2025recommender,zhao2024toolrec,yang2025agentdr,park2025madrec,ma2025agenticrag,chen2025vragent,gu2025r}. These metrics should be interpreted with ablations. In L2 systems, gains may arise from better retrieval or prompt construction; in L3 systems, they may hide brittle tool routing; in L4 systems, they may depend on reflection prompts; and in L5 systems, they may reflect one useful specialist rather than effective coordination. Reporting should therefore include the candidate set, negative sampling protocol, split, baselines, LLM version, number of tool calls, memory size, reflection rounds, and repeated-run variance.

Cold-start and sparse-data settings illustrate why classical metrics remain important. In these settings, agentic components may supply missing preference evidence, convert natural-language constraints into structured signals, or generate synthetic interactions. Ranking metrics can measure whether these additions improve downstream recommendation utility. They do not, however, establish why the improvement occurs.

Conversational ARS add an interaction layer between intent and recommendation. Evaluation must measure not only whether a target item appears, but also whether the agent elicits missing preferences, follows commands, adapts to feedback, remains coherent across turns, and stops at an appropriate time. Common metrics include success rate, Success@K, average turns, task completion, satisfaction, engagement, and willingness to reuse. MACRS and ECPO are representative because they evaluate multi-turn conversational recommendation and expectation confirmation rather than only static ranking \cite{fang2024multi,feng2025expectation}. RecBot, SmartEats, WeMusic-Agent, and AdaptJobRec further illustrate settings where command following, dietary constraints, music interaction, or job-preference elicitation make user experience part of the recommendation target \cite{tang2025interactive,liang2025smarteats,wang2025adaptjobrec, haller2026impress}.

Explanations and other generated outputs require separate evaluation. A rationale can be fluent but unsupported, concise but unhelpful, or faithful to an item attribute but irrelevant to the user's current constraint. MADREC and REXHA evaluate explanation quality, CARTS focuses on recommendation textual summarization, and AgenticRAG combines recommendation with retrieval-grounded explanation \cite{park2025madrec,sun2025retrieval,chen2025carts,ma2025agenticrag}. Automatic text-overlap metrics such as BLEU, ROUGE, METEOR, and BERTScore are useful when references exist, but they do not establish factual support or user usefulness. Human or LLM judges should therefore score at least two axes: communicative quality (clarity, relevance, naturalness, personalization, usefulness) and evidence faithfulness (whether factual claims are supported by item metadata, user history, reviews, knowledge-base triples, or retrieved passages). Judge prompts, rubrics, visibility of evidence, identity masking, item-order randomization, and agreement statistics should be reported.

\subsection{Grounded Retrieval and RAG Evaluation}
\label{subsec:rag-eval}

Retrieval-augmented generation is common in ARS because recommendation often depends on information outside model parameters: user profiles and histories, item attributes, reviews, knowledge graphs, web pages, policy rules, and domain constraints. In the LoA taxonomy, this is central to L2 systems and becomes a component of L3--L5 systems when retrieval is embedded in tools, planning, or multi-agent workflows. RAG evaluation should distinguish retrieval quality from grounded generation. A recommender can retrieve relevant items but generate an unsupported explanation, or retrieve weak evidence and still guess a plausible item.

The evidence space has three sources:  \emph{user information}, \emph{item information}, and \emph{auxiliary knowledge}. 
RA-Rec, CORAL, ARAG, and AgenticRAG illustrate retrieval-augmented recommendation and explanation settings where evidence quality should be checked separately from final ranking \cite{kemper2024retrieval,wu2024coral,maragheh2025arag,ma2025agenticrag}. 
1) \textbf{Retriever diagnostics} should include Recall@K, Precision@K, MRR, NDCG, coverage of required attributes, diversity of evidence sources, and freshness when data changes over time. 
2) \textbf{Generator diagnostics} should include answer relevance, claim-level citation accuracy, faithfulness, hallucination rate, contradiction rate, and consistency with the dialogue context. 
3) \textbf{Robustness tests} should include insufficient evidence, conflicting evidence, poisoned item descriptions, and unsupported user requests.

RAG evaluation should also report how the retrieved context is used. Merely retrieving a relevant review or item attribute does not guarantee that the generator used it correctly. Claim-level checking is preferable for user-facing rationales: each factual statement should be linked to a supporting user profile entry, item attribute, review, knowledge-graph fact, policy rule, or web/API result. When the system compares alternatives, the comparison should be checked against the same candidate set and the same time-sensitive values, such as price, availability, or distance. Negative controls are also useful: if evidence is removed or contradicted, the agent should either abstain, ask a clarification, or revise the claim rather than generate unsupported text.

\subsection{Simulator Evaluation}
\label{subsec:simulator-eval}

Simulation is both a method for evaluating ARS and an object that must be evaluated~\cite{yoon2024evaluating, zhang2026exploring}. A simulator that improves downstream NDCG can still misrepresent real users, reveal labels too directly, collapse behavioral diversity, or reward policies that exploit simulator artifacts. Simulator evaluation should therefore distinguish \emph{downstream utility} from \emph{simulator validity}. 
1) \textbf{Downstream utility} asks whether simulated interactions improve a recommender under HR, NDCG, Recall, reward, liking ratio, or inference-time metrics. Validity asks whether simulated users, creators, and environments behave like the intended population. 
2) \textbf{Validity checks} should include distributional fidelity to logs, behavioral realism under persona constraints, diversity across personas, leakage of target items or labels, and calibration against human judgments or online treatment effects. Long-term environments should additionally report preference drift, novelty decay, provider exposure, diversity, retention, and feedback-loop effects. Without such checks, simulation may be useful for development but weak as evidence of real-world behavior.

A useful simulator report should make the validity conditions explicit. If the simulator is calibrated on a particular domain, interaction type, or user population, the paper should not imply that its conclusions transfer automatically to other domains. It should also state what the simulator observes. A simulator with access to hidden target labels, full item metadata, or future interactions can produce overly optimistic results. A simulator that produces both implicit feedback and rationales should evaluate the two outputs separately, because plausible text does not imply realistic click or rating behavior. For long-horizon environments, the report should include stability checks over repeated interactions; otherwise small response-model errors may compound into misleading claims about retention, fairness, or filter bubbles.

\subsection{Trace, Deployment, and Trustworthiness Evaluation}
\label{subsec:trace-deployment-trust-eval}

Agentic systems expose intermediate traces that should be evaluated directly. 
1) For tool-driven agents, the trace should show whether the agent select the correct tool, supplied valid arguments, interpret the result correctly, recover from tool failures, and avoid unnecessary calls. ToolRec, RecMind, iAgent, AgentDR, and TAIRA motivate such diagnostics because their claimed contribution depends on explicit tool use, planning, shielding, or thought-pattern selection \cite{zhao2024toolrec,wang2024recmind,xu2025iagent,yang2025agentdr,yu2025thought}. For memory, evaluation should measure retrieval precision and recall, update faithfulness, stale-memory use, contradiction rate, and privacy leakage. 
2) For planning, evaluation should check task decomposition, constraint coverage, step ordering, verification, and recovery. For multi-agent systems such as MACRS, MACRec, Collab-REC, MAP, and MATCHA, evaluation should include agreement, vote entropy, number of rounds, time-to-consensus, role ablations, deadlock frequency, and the marginal utility of each specialist agent \cite{fang2024multi,wang2024macrec,banerjee2025collab,lee2025map,hui2025matcha}.

Deployment evaluation is where many agentic designs become difficult to justify. Customer-facing ARS need latency, token cost, inference cost, throughput, tool-call count, cache-hit rate, timeout rate, fallback rate, and reliability metrics, in addition to online outcomes such as CTR, CVR, retention, satisfaction, or GMV. ColdLLM is an example of combining offline cold-start metrics with online-style business evidence, while efficiency-aware RAG and cloud-device collaborative recommendation show that accuracy must be studied jointly with latency, cost, and system boundaries \cite{huang2025large,zhou2025efficiency,long2025cloud}. The most informative reports present trade-offs, such as NDCG versus latency or success rate versus tokens per episode, rather than a single best score. Parallelism should also be reported, since multi-agent systems may reduce wall-clock time while increasing total token cost and orchestration complexity.

Deployment metrics should be reported with the same care as ranking metrics. Mean latency is insufficient without tail latency, since multi-step agents may fail through long-tail tool delays. Token cost should distinguish prompt, retrieval, reasoning, verifier, and explanation stages where possible. Tool-call count should be reported per episode and, for multi-agent systems, per role. Fallback behavior should also be measured: a system that silently falls back to a generic ranker after tool failure may appear reliable but no longer exercises the agentic workflow being studied. Online experiments should state traffic split, duration, confidence intervals, guardrail metrics, and whether users were exposed to generated explanations or only to final items.

Trustworthiness should be a primary evaluation target. RAG exposes item descriptions and knowledge bases to poisoning; memory can be corrupted, stale, or privacy-sensitive; tool use can trigger unsafe API calls; and multi-agent communication can propagate inconsistent state. Robustness protocols should include adversarial item descriptions, shilling or poisoning attacks, prompt injection, memory perturbation, and corrupted retrieval evidence. The poisoning study on retrieval-augmented recommenders and the DrunkAgent memory-perturbation work illustrate these risks \cite{nazary2025stealthy,yang2025get,ning2024cheatagent, li2026agentattack, gu2026llm}. Privacy evaluation should measure what is stored, where it is stored, how it is retrieved, whether it can be deleted, and whether sensitive attributes are exposed to other agents or tools \cite{zhang2025towards,long2025cloud}. Fairness and diversity evaluation should consider users, items, creators, and providers; multi-stakeholder recommendation and long-term simulation show that exposure and diversity effects may appear only after repeated interaction \cite{banerjee2025collab,jannach2025rethinking,ye2025creagent,sukiennik2025simulating}.

\begin{table*}[!t]
\centering
\footnotesize
\setlength{\tabcolsep}{4pt}
\renewcommand{\arraystretch}{1.08}
\caption{Minimum reporting checklist for ARS evaluation.}
\vspace{-8pt}
\label{tab:ars-reporting-checklist}
\begin{tabularx}{\textwidth}{@{}p{0.17\textwidth}L@{}}
\toprule
\textbf{Evaluation component} & \textbf{Minimum information to report} \\
\midrule
Offline recommendation & Data split, candidate set, negative sampling, baselines, repeated runs, LLM/backbone version. \\
Conversation and generation & User or simulator protocol, task definition, stopping criteria, judge rubric, prompt, evidence visibility, agreement statistics. \\
RAG and grounding & Retrieval source, evidence granularity, freshness, retrieval budget, claim segmentation, faithfulness or contradiction checks. \\
Tools and planning & Tool schema, tool-call logs, argument-validity checks, execution success, recovery policy, planner or verifier ablation. \\
Memory & Memory type, update rule, retrieval budget, stale-memory policy, contradiction checks, privacy and deletion handling. \\
Multi-agent coordination & Agent roles, message format, number of rounds, consensus rule, deadlock handling, role ablations, cost per role. \\
Safety and privacy & Threat model, attack budget, policy test suite, defense baselines, sensitive attributes, leakage metric. \\
Deployment & Hardware/API setup, latency distribution, token and inference cost, throughput, timeout/fallback rates, traffic split, confidence intervals. \\
\bottomrule
\vspace{-15pt}
\end{tabularx}
\end{table*}

A practical evaluation stack should therefore be layered. Offline metrics provide fast regression tests for relevance. Trace metrics diagnose tools, memory, plans, and coordination. Grounding metrics audit factual support. Simulator metrics test controlled interaction and long-horizon effects, subject to calibration. Human or LLM judges assess aspects of communication that are difficult to reduce to item relevance. Deployment metrics test whether the design is feasible under latency, cost, and reliability constraints. Online experiments, when available, provide the strongest evidence for user and business impact. No single protocol is sufficient for all ARS claims.

The checklist should be read as a minimum rather than a complete benchmark specification. Different applications may need additional domain-specific measurements. A travel planner may require route feasibility and weather or availability checks; a food recommender may require dietary and allergen constraints; a learning-path recommender may require prerequisite consistency; and a marketplace recommender may require provider exposure and inventory constraints. These domain-specific checks should be attached to the relevant target in Table~\ref{tab:ars-eval-matrix} rather than introduced as isolated metrics. This keeps the evaluation design interpretable: the reader can see which system claim is being tested, which protocol supports it, and which metric is used to report it.

\subsection{Open Evaluation Gaps}
\label{subsec:evaluation-future}

The main evaluation gap is that architecture has advanced faster than measurement. Many papers introduce memory, tools, RAG, planning, reflection, or multiple agents, but still evaluate mainly with static ranking metrics. Future benchmarks should record full trajectories and compute layered metrics for recommendation utility, interaction quality, generated-output quality, grounding, tool correctness, memory behavior, safety, and cost. Calibrated simulation is another priority: simulator papers should report distributional fidelity, behavioral probes, leakage controls, diversity, uncertainty, and correlation with human or online outcomes. Finally, multi-agent evaluation should make coordination measurable through message-level diagnostics, specialist-agent ablations, contribution analysis, and cost-latency-quality trade-offs. These practices would make it clearer when agentic components improve recommendation and when a simpler agent-assisted or conventional pipeline is sufficient.

\section{Open Challenges and Opportunities}

Agentic recommender systems introduce new capabilities beyond traditional recommendation pipelines, yet they also expose several unresolved challenges. We organize the open problems and opportunities from three complementary perspectives: the modeling perspective (Section~\ref{subsec:challenge_modeling}), the user perspective (Section~\ref{subsec:challenge_user}), and the system perspective (Section~\ref{subsec:challenge_system}).

\subsection{Modeling Perspective}\label{subsec:challenge_modeling}

\subsubsection{Lifelong User Modeling}
A fundamental challenge lies in capturing a user's evolving interests across long time horizons. Existing agentic systems often treat user representations as short-term context, which limits their ability to retain stable preferences while adapting to new signals. Agents may overfit to recent interactions or lose information about inactive users.
An opportunity is to design lifelong memory mechanisms that distinguish persistent preferences from transient behaviors. This can be operationalized through hierarchical memory blocks, user-specific episodic buffers, or retrieval-augmented preference summaries. Another direction is to define a new evaluation task on lifelong preference retention, allowing systematic assessment of how well an agent maintains user identity across continual updates.

\subsubsection{Contextual Modeling}
Recommendation relevance depends strongly on context, yet most current agent workflows do not differentiate between short-term triggers, mid-term sessions, and long-term routines. Agents often rely on raw history, which mixes signals of varying importance and introduces noise.
A promising opportunity is to adopt context engineering principles that help the agent identify, compress, and structure contextual information before reasoning. This includes context abstraction modules that re-write user history into semantically meaningful summaries, as well as context-selection tasks where the agent must choose an optimal subset of history for the downstream recommendation. Such mechanisms encourage structured reasoning and reduce reliance on long unfiltered sequences.

\subsubsection{Multimodal Modeling}
Modern recommendation scenarios incorporate text, images, video, and audio. Agents face two major challenges: building unified multimodal representations and aligning these signals with user preferences. Naïve fusion often leads to inconsistent semantics and weak cross-modal grounding.
Opportunities include designing multimodal tokenization schemes that convert heterogeneous content into a shared token space, enabling smoother reasoning. Cross-modal alignment modules can be integrated into the agent workflow to enforce consistency between modalities. In addition, multimodal reasoning tasks can be formulated to benchmark how well the agent interprets visual and textual cues jointly. These tasks can guide the development of agents that can explain recommendations grounded in both semantics and appearance.

\subsection{User Perspective}\label{subsec:challenge_user}

\subsubsection{Controllability}
Users often lack clarity on how an agent arrives at a decision. Current agentic systems offer limited control over the decision process or the ability to steer recommendations through explicit feedback.
Opportunities arise in designing interfaces and agent workflows that expose intermediate decisions. Decision-process visibility can be supported by traceable reasoning graphs. Feedback-based control modules can allow users to adjust preference dimensions or veto specific reasoning paths. This improves both personalization and system transparency.

\subsubsection{Explainability and Trustworthiness}
Agentic recommender systems rely on multi-step reasoning, but these chains may introduce errors, hallucinations, or unsupported claims. Users may find it difficult to trust an opaque reasoning process.
Agents can incorporate self-verification modules that check consistency across reasoning steps. Structured explanation templates can enforce factual grounding and avoid speculative statements. Defining trustworthiness benchmarks for agentic recommendation—for instance, accuracy of reasoning chains or stability across perturbations—would help standardize evaluation and guide future development.

\subsubsection{Privacy Preservation}
Agents operate on large contextual windows and may inadvertently expose sensitive information during reasoning or tool use. Maintaining user privacy across memory, retrieval, and agent-to-agent communication is challenging.
Opportunities include privacy-aware memory architectures that control what information is stored or exposed, and fine-grained permission protocols that regulate tool usage. Developing privacy-preserving reasoning tasks would further encourage models to separate preference inference from identifiable personal details.

\subsection{System Perspective}\label{subsec:challenge_system}

\subsubsection{Scalability}
Agentic systems often require multiple agents or repeated workflow iterations to complete a task, which leads to scalability issues in large-scale recommendation environments. It remains unclear when specialized agents are necessary or how to coordinate multi-agent collaboration.
A solution is to introduce agent orchestration frameworks that determine when to call which agent, based on task complexity or user state. Designing adaptive workflow planners that limit unnecessary reasoning cycles can further improve scalability. A benchmark for multi-agent recommendation settings would also help identify the optimal structure for different scenarios.

\subsubsection{Efficiency in Training and Inference}
Training and serving agentic models can be computationally expensive due to long context windows, multi-step reasoning, and external tool calls. Inference latencies may restrict real-world adoption.
Opportunities lie in designing efficient training paradigms, such as distilling multi-step reasoning into shorter latent plans, storing reusable intermediate states, or learning compact preference embeddings. For inference, caching partial reasoning results and developing fast verification modules can significantly reduce latency. Clear system-level metrics that capture efficiency–performance trade-offs will guide deployment decisions.

\subsubsection{API and Interaction-Level Efficiency}
Many agentic systems rely on external APIs for search, item retrieval, or environment interaction. Excessive calls can bottleneck performance and increase costs.
One opportunity is to develop local proxy tools or lightweight internal knowledge stores that reduce dependency on external APIs. Another is to design agent–tool protocols that batch operations or estimate when tool calls are unnecessary. These improvements can stabilize real-time recommendation workloads.
\section{Conclusions}
Agentic recommender systems represent a significant shift from passive item-ranking pipelines to autonomous, goal-directed systems capable of reasoning, planning, acting, and adapting within complex environments. 
By integrating large language model agents with profiles, memory structures, tool interfaces, and multi-step workflows, modern systems move beyond traditional assumptions about static user intents and fixed recommendation stages. Our survey organizes this emerging field through three complementary paradigms, \ie agent-assisted recommendation, agent-as-recommender, and agent-as-user-simulator. This survey situates them within a unified autonomy framework that spans retrieval augmentation, tool use, single-agent planning, and multi-agent collaboration.

Across studies, we observe clear trends. Agent-assisted systems demonstrate strong gains in semantic understanding, retrieval quality, and contextual relevance. Single-agent recommenders further introduce coherent reasoning, long-horizon planning, and dynamic preference modeling, while multi-agent frameworks enable division of labor, critique, negotiation, and consensus building. User simulation agents provide new opportunities for scalable evaluation, counterfactual analysis, and environment modeling, allowing the field to explore settings that are difficult to observe in real-world logs. 
Despite these advances, substantial challenges remain. Lifelong user modeling, contextual abstraction, and multimodal grounding are still far from solved. User experience issues such as controllability, explainability, trustworthiness, and privacy require dedicated attention as systems become more autonomous. At the system level, scalability, efficiency, and robust agent–tool orchestration will determine whether agentic approaches can be deployed at industrial scale. Addressing these challenges creates opportunities to propose new tasks, design specialized modules, and introduce capabilities that reshape the user experience of recommendation.

\bibliography{bibtex}

\end{document}